\documentclass[12pt,a4paper,oneside]{article}

\usepackage[margin=2cm]{geometry}

\usepackage{amsmath}
\usepackage{graphicx}
\usepackage{dcolumn}
\usepackage{bm}
\usepackage{bbm}
\usepackage{epsfig}
\usepackage{slashed}
\usepackage{amssymb}

\newcommand{\newc}{\newcommand}
\newc{\renewc}{\renewcommand}

\def\beq{\begin{equation}}
\def\eeq{\end{equation}}
\def\bea{\begin{eqnarray}}
\def\eea{\end{eqnarray}}
\def\bitem{\begin{itemize}}
\def\eitem{\end{itemize}}
\def\ba{\begin{array}}
\def\ea{\end{array}}
\def\bal{\begin{align}}
\def\eal{\end{align}}
\def\bi{\begin{itemize}}
\def\ei{\end{itemize}}
\def\lsim{\mathrel{\rlap{\lower4pt\hbox{\hskip1pt$\sim$}}
    \raise1pt\hbox{$<$}}}         
\def\gsim{\mathrel{\rlap{\lower4pt\hbox{\hskip1pt$\sim$}}
    \raise1pt\hbox{$>$}}}

\newc{\ie}{{\it i.e.~}}          \newc{\etal}{{\it et al.~}}
\newc{\eg}{{\it e.g.~}}          \newc{\etc}{{\it etc.~}}
\newc{\cf}{{\it c.f.~}}
\newc{\os}{\mbox{\hspace{4pt}}}
\newc{\us}{\mbox{\hspace{12pt}}}

\renewc{\bar}{\overline}

\newc{\gev}{\,{\rm GeV}}
\newc{\mev}{\,{\rm MeV}}
\newc{\ev}{\,{\rm eV}}
\newc{\kev}{\,{\rm keV}}
\newc{\tev}{\,{\rm TeV}}

\newc{\LM}{\mathcal{L}}
\newc{\SM}{\mathcal{S}}

\newc{\HM}{\mathcal{H}}
\newc{\GM}{\mathcal{G}}
\newc{\OM}{\mathcal{O}}
\newc{\FM}{\mathcal{F}}
\newc{\AM}{\mathcal{A}}
\newc{\BM}{\mathcal{B}}
\newc{\NM}{\mathcal{N}}
\newc{\WM}{\mathcal{W}}
\newc{\ZM}{\mathcal{Z}}

\newc{\Chi}{\mathcal{X}}

\begin{document}

\title{ {\tiny \vspace*{-2.cm}\hspace*{14cm} \phantom{*}}\\
{\tiny \vspace*{-.5cm} \hspace*{14cm}  \phantom{*}} \vspace*{.9cm}\\ 
\bf \Large Constraints on composite quark partners\\
\bf \Large from Higgs searches}

\author{\fontsize{12}{16}\selectfont Thomas Flacke$^{\, a}$, Jeong Han Kim$^{\, a,b}$, Seung J. Lee$^{\, a,c}$ and Sung Hak Lim$^{\, a,b}$
\vspace{6pt}\\
\fontsize{11}{16}\selectfont\textit{$^a$ Department of Physics, Korea Advanced Institute of Science and Technology,}\\
\fontsize{11}{16}\selectfont\textit{335 Gwahak-ro, Yuseong-gu, Daejeon 305-701, Korea}\\
\fontsize{11}{16}\selectfont\textit{$^b$ Center for Theoretical Physics of the Universe, IBS, Daejeon, Korea}\\
\fontsize{11}{16}\selectfont\textit{$^c$ School of Physics, Korea Institute for Advanced Study, Seoul 130-722, Korea}} 
\date{}
\maketitle

\begin{abstract}
In composite Higgs models, the generation of quark masses requires the standard model-like quarks to be partially or fully composite states which are accompanied by composite quark partners. The composite quark partners decay into a standard model-like quark and an electroweak gauge boson or Higgs boson, which can be searched for at the LHC.
In this article, we study the phenomenological implications of composite quarks in the minimal composite Higgs model based on the coset $SO(5)/SO(4)$. We focus on light quark partners which are embedded in the $SO(4)$ singlet representation. In this case, a dominant decay mode of the partner quark is into a Higgs boson and a jet, for which no experimental bounds have been established so far. The presence of $SO(4)$ singlet partners leads to an enhancement of the di-Higgs production cross section at the LHC. This will be an interesting experimental signature in the near future, but, unfortunately, there are no direct bounds available yet from the experimental analyses.
However, we find that the currently available standard model-like Higgs searches can be used in order to obtain the first constraints on partially and fully composite quark models with light quark partners in the $SO(4)$ singlet. We obtain a flavor- and composition parameter independent bound on the quark partner mass of $M_{U_h} > 310 \gev$ for partially composite quark models and  $M_{U_h} > 212 \gev$ for fully composite quark models. 
\end{abstract}

\newpage



\section{Introduction}

The recent discovery of a Higgs-like boson at the LHC~\cite{Aad:2012tfa,Chatrchyan:2012ufa} represents a remarkable success for the Standard Model (SM) of particle physics. However, within the SM the Higgs mass is subject to additive renormalization, implying that a large hierarchy between the electroweak scale and the Planck scale is technically unnatural~\cite{thooft}.
One of few motivated models beyond the Standard Model (BSM) addressing this so-called fine-tuning problem is the framework of composite Higgs models~\cite{HPGB,Agashe:2004rs} with the Higgs as a pseudo-Nambu-Goldstone-boson associated with the spontaneous breakdown of an approximate global symmetry of a sector which becomes strongly coupled at a scale $f$. The global symmetry is explicitly broken by Yukawa couplings of the Higgs to the quarks and their composite partner states. Integrating out the quark partners yields an effective potential for the Higgs. The effective potential strongly depends on the embedding of the top quark partners into the global symmetry group as well as the top partner mass, while the lighter quark partners typically play a minor role. Concrete realizations of composite Higgs models \cite{Matsedonskyi:2012ym,Redi:2012ha,Marzocca:2012zn,Pomarol:2012qf,Panico:2012uw}, electroweak precision constraints \cite{Grojean:2013qca, Gillioz:2013pba}, and top-partner phenomenology \cite{DeSimone:2012fs, Buchkremer:2013bha,CMStoppartners,Vignaroli:2012nf,Azatov:2013hya} have already been studied, while only few articles focussed on bottom partners \cite{Gillioz:2013pba,Vignaroli:2012sf} or partners of other light quarks \cite{CHus}.

Although typically not majorly contributing to electroweak symmetry breaking, light quarks need to be accompanied by composite partners in order to generate (small but non-vanishing) masses of the SM-like light quarks.\footnote{
Often, the composite sector is assumed to be flavor-blind in order to avoid constraints from flavor changing neutral currents (\cf \eg Ref.~\cite{Redi-Weiler}).
Such a choice would imply the light quark partners are degenerate with the top quark partners, up to Yukawa-suppressed corrections. However, as has been pointed out in \cite{Gedalia-Kamenik-Ligeti-Perez}, partners are allowed to be non-degenerate within models of flavor alignment~\cite{Cacciapaglia:2007fw,Fitzpatrick-Perez-Randall, Csaki-Perez-Surujon-Weiler}.
It was also shown that while electroweak precision tests put severe constraints on the degree of compositeness of the SM quark doublets~\cite{flavor-triviality,Redi-Weiler},
due to an approximate custodial parity~\cite{Zbb}, the bounds can be much weaker for the SM quark singlets.
In this article we therefore allow for non-degenerate quark partner masses and treat them as free parameters.}
The quark partner phenomenology depends on the embedding of the quarks and their partners into the global symmetry group of the composite sector, as well as their masses and couplings. In Ref.~\cite{CHus}, quark partners of the up and charm quarks were studied within the minimal composite Higgs model based on $SO(5)/SO(4)$. The right-handed quark partners were embedded in the $\bf 5$ of $SO(5)$, which comprises of a $\bf 4$ and a singlet under the $SO(4)$ which is unbroken by strong dynamics. While bounds for the partners in the $\bf 4$ were obtained, the singlet partner remained unconstrained because of its suppressed couplings to electroweak gauge bosons. It was furthermore shown, that the presence of a light singlet state can substantially weaken the constraints on the partners in the fourplet, due to a combination of smaller production rates of fourplet states, and the opening of cascade decay processes of fourplet states via the singlet state.

In this article, we focus on constraints of quark partners in the $SO(4)$ singlet in composite Higgs models based on the minimal coset $SO(5)/SO(4)$. In Sec.~\ref{sec:models} we discuss the two basic phenomenologically viable setups for composite quarks in the $SO(4)$ singlet, where the right-handed quark is either realized as an elementary quark which mixes with a composite partner (partial compositeness) or as a chiral state of the composite sector (full compositeness).  We establish effective Lagrangians for both these models and use those in order to discuss their LHC phenomenology in In Sec.~\ref{sec:pheno}. In particular we show that both setups  can result in a substantial increase of di-Higgs production above the SM background if the quark partners are light. In the current absence of direct bounds on the di-Higgs production channel at the LHC, we use the ATLAS bounds on differential cross sections of the Higgs di-photon decay in order to obtain constraints on the partially and fully composite light quark models in Sec.~\ref{sec:results}, where we present our results in terms of the effective models discussed in Sec.~\ref{sec:models}. We conclude in Sec.~\ref{sec:conclusions}.


\section{Models}\label{sec:models}

As a generic setup, we consider the fermion sector of the minimal composite Higgs model ~\cite{Agashe:2004rs} based on the coset structure $SO(5)$$/$$SO(4)$. We follow the conventions and notation of Ref.~\cite{CHus} and use the setups presented there for an initial discussion. We start the discussion with the embedding of up-type partners. For concreteness, we embed the left-handed elementary quark doublet $q_L$ in an incomplete $\bf 5$ of $SO(5)$~\cite{Contino:2006qr,CHus} as
\beq
\bar{q}_L^{5}=\frac{1}{\sqrt{2}}\left(-i\bar{d}_L\, , \bar{d}_L\, , -i\bar{u}_L\, , -\bar{u}_L\, , 0\right).
\eeq
For the up-type partners, the $q_L$ carries a $U(1)_X$ charge of $2/3$. 
The $U(1)_X$ is included in order to generate the correct hyper charge of the quarks, which is obtained by gauging $Y=T^3_R+X$, where $T^3_R$ is the diagonal generator of the $SU(2)_R$ in $SO(4)\simeq SU(2)_L\times SU(2)_R$.

The fermionic partners of the up-type quarks are also included in a $\bf 5$ of $SO(5)$ (with $U(1)_X$ charge $2/3$) as
\beq
\psi=\left(\begin{array}{c}
                Q\\
		  \tilde{U}
               \end{array}\right)
               =\frac{1}{\sqrt{2}}\left(\begin{array}{c}
                iD-iX_{5/3}\\
		 D+X_{5/3}\\
		 iU+iX_{2/3}\\
		 -U+X_{2/3}\\
		 \sqrt{2}\tilde{U}
               \end{array}\right)
               \underrightarrow{\, \, M_4\rightarrow\infty \, \, }
               \left(\begin{array}{c}
               0\\
		0\\
		0\\
		0\\
		\tilde{U}
               \end{array}\right),        
\label{defpsi}            
\eeq
which can be decomposed as a fourplet, $Q$, with a mass scale $M_4$ and a singlet, $\tilde U$, with a mass scale $M_1$ of the $SO(4)$ which is unbroken by the strong dynamics. In this article, we are studying the singlet partner and therefore take the limit $M_4\rightarrow \infty$ in which the fourplet partners decouple, while the singlet partner $\tilde{U}$ remains as the only BSM particle.

The interactions of the $\tilde{U}$ depend on the embedding of the right-handed quarks. They can be either embedded as a chiral composite $SO(5)$ singlet $u_R$ (in ``fully composite'' models)~\cite{Pomarol:2008bh} or as an incomplete  representation of $SO(5)$ (in ``partially composite" models) which is a singlet state in terms of the $SO(4)$. For concreteness, we embed $u_R$ into the $\bf 5$ of $SO(5)$ as
\beq
\bar{u}_R^5=\left(0,0,0,0,\bar{u}_R\right).
\eeq
Such an embedding is termed partial compositeness, because the mass eigenstate is a linear combination of the ``elementary'' quarks and the composite partner $\tilde{U}$. In both -- fully and partially composite -- embeddings, the right-handed quarks have a $U(1)_X$ charge of $2/3$. 
Down-type partners can be embedded analogously with a  $U(1)_X$ charge of $-1/3$ for $q^5_L$, $d_R$ and $\psi^d$.\footnote{For down-type partners, the embeddings of the $q_L$ and $\psi^d$ read $\bar{q}^5_L=\left(1/\sqrt{2}\right)\left(i\bar{u}_L\, , \bar{u}_L\, , -i\bar{d}_L\, , \bar{d}_L\, , 0\right)$ and $\bar{\psi}^d=\left(1/\sqrt{2}\right)\left( i\bar{U}-i\bar{X}_{-4/3},\bar{U}+\bar{X}_{-4/3},-i\bar{D}-i\bar{X}_{-1/3},-\bar{D}+ \bar{X}_{-1/3},\sqrt{2} \bar{\tilde{D}} \right)$ .}

The above embedding is not unique for composite Higgs models based on $SO(5)$$/$$SO(4)$. Other embeddings of left-handed and partner quarks discussed in the literature include the symmetric $\bold{14}$ representation of $SO(5)$ \cite{Pomarol:2012qf,Panico:2012uw,Pappadopulo:2013vca,Montull:2013mla}.\footnote{Refs.~\cite{Contino:2006qr, Azatov:2011qy} also discuss embeddings into the anti-symmetric $\bold{10}$ representation. However, the $\bf 10 $ decomposes into $\bf{6} \oplus \bf{4}$ in terms of $SO(4)$ representations and does not contain an $SO(4)$ singlet partner as is considered, here.} However, we will be focussing on the $SO(4)$ singlet partner $\tilde{U}$ (or the equivalent partner for down-type quarks). Our results are to a large part independent of the $SO(5)$ representation in which the quarks are embedded in, as long as it contains the $SO(4)$ singlet. In the following, we derive an effective description for the $SO(4)$ singlet, using partners in the $\bold{5}$ as a guideline and comment on how this can be applied to other representations of quark partners.   


\subsection{Partially composite quark models}\label{sec:PCtheory}

We start with the partially composite model outlined in Ref.~\cite{CHus}. Using the Callan-Coleman-Wess-Zumino (CCWZ) formalism~\cite{ccwz}, the fermion Lagrangian of the partially composite model is
\beq
\mathcal{L}=\mathcal{L}_{\rm comp}+\mathcal{L}_{\rm el, mix}
\eeq
with
\beq
\mathcal{L}_{\rm comp}=i\ \bar{Q}(D_\mu +i e_\mu) \gamma^\mu Q + i \bar{\tilde U}\slashed{D}\tilde U
 -M_4\bar{Q}Q
 -M_1\bar{\tilde{U}}\tilde{U}+\left(i c \bar{Q}^i \gamma^\mu d^i_\mu \tilde{U}+\mbox{h.c.}\right),
\eeq
where  $e_\mu$ and $d^i_\mu$ are the CCWZ connections (\cf Appendix~A of Ref.~\cite{CHus} for the explicit expressions), and 
\beq
 \mathcal{L}_{el,mix}=i\ \bar{q}_L\,\slashed{D}q_L+i\ \bar{u}_R\,\slashed{D}u_R-y_L f\bar{q}^5_L \mathcal{F}(U_{\rm gs})\psi_R
 -y_R f\bar{\psi}_L \mathcal{G}(U_{\rm gs})u_R+\mbox{h.c.},\label{Lelmixpc}
\eeq
where $\mathcal{F}$ and $\mathcal{G}$ are functions of the Goldstone matrix 
\beq
U_{\rm gs}=\left(\begin{array}{ccccc}
                1 & 0 & 0 & 0 & 0 \\
		 0 & 1 & 0 & 0 & 0 \\
		 0 & 0 & 1 & 0 & 0 \\
		 0 & 0 & 0 & \cos{\frac{h+v}{f}} & \sin{\frac{h+v}{f}} \\
		 0 & 0 & 0 & -\sin{\frac{h+v}{f}} & \cos{\frac{h+v}{f}}
               \end{array}\right)\,,
               \label{gmatrU}
\eeq
with appropriate $SO(5)$ index contractions such that the action is $SO(5)$ invariant. For $u_L$, $\psi$ and $u_R$ in the $\bold{5}$ these are simply $\mathcal{F}=\mathcal{G}=U^{i5}_{\rm gs}$. In the limit $M_4\rightarrow \infty$ in which the $SO(4)$ fourplet states $Q$ decouple, one obtains
\bea
\mathcal{L}&=& i \bar{\tilde U}\slashed{D}\tilde U -M_1\bar{\tilde{U}}\tilde{U} +  i\ \bar{q}_L\,\slashed{D}q_L +i\ \bar{u}_R\,\slashed{D}u_R\nonumber\\
&&-\left[\frac{y_L}{\sqrt{2}} f\bar{u}_L F\left(\frac{h+v}{f}\right)\tilde{U}_R
 +y_R f\bar{\tilde{U}}_L G\left(\frac{h+v}{f}\right)u_R+\mbox{h.c.}\right]\label{pcLag1}\\
&=&  i \bar{\tilde U}\slashed{D}\tilde U -M_1\bar{\tilde{U}}\tilde{U} +  i\ \bar{q}_L\,\slashed{D}q_L +i\ \bar{u}_R\,\slashed{D}u_R\nonumber\\
&&-\left[m_2  \bar{u}_L \tilde{U}_R+m_3  \bar{\tilde{U}}_L u_R+\lambda_2 h \bar{u}_L \tilde{U}_R+\lambda_3 h \bar{\tilde{U}}_L u_R+  \mbox{h.c.} +\mathcal{O}(\epsilon^2)\right].\nonumber
\eea
In the above, the functions $F$ and $G$ are $F=-\sin((h+v)/f)$ and $G=\cos((h+v)/f)$ for $u_L$, $\psi$ and $u_R$ in the $\bold{5}$. In order to obtain the final expression, we expanded the Lagrangian in $\epsilon=v/f$ which yields
\bea
m_{2}&=& \frac{y_L}{\sqrt{2}} f F(\epsilon) = -\frac{y_L}{\sqrt{2}} f \sin(\epsilon),\nonumber\\
m_3&=& y_R f G(\epsilon) = y_R f \cos(\epsilon),\nonumber\\
\lambda_2&=&  \frac{y_L}{\sqrt{2}}  F'(\epsilon) = - \frac{y_L}{\sqrt{2}} \cos(\epsilon), \nonumber\\
\lambda_3&=&  y_R  G'(\epsilon) = -y_R \sin(\epsilon).
\eea

From Eq.~\eqref{pcLag1} we read off the effective fermion mass terms
\beq
\mathcal{L}_m=-(\bar{u}_L, \bar{\tilde{U}}_L) M_u \left(\begin{array}{c} u_R\\ \tilde{U}_R \end{array}\right)+ \mbox{h.c.}
\mbox{\hspace{20pt} with \hspace{20pt}}
M_u= \left(
\begin{array}{cc}
0& m_2\\
m_3 & M_1
\end{array}
\right).
\eeq
Diagonalizing the mass matrix yields the mass squared eigenvalues
\bea
M^2_{u_l,U_h}&=&\frac{1}{2}\left[\left(M_1^2+m_2^2+m_3^2\right)\mp \sqrt{\left(M_1^2+m_2^2+m_3^2\right)^2-4 m^2_2 m_3^2}\right]\nonumber\\
&=&\frac{1}{2}\left(M_1^2+m_2^2+m_3^2\right)\left[1\mp\sqrt{1-\frac{4m^2_2 m_3^2}{\left(M_1^2+m_2^2+m_3^2\right)^2}}\right].
\eea
Requiring that the lighter eigenvalue reproduces the quark mass square of the SM quarks which is (much) smaller than $M_1$ for (1st and 2nd) 3rd family quarks thus requires the last term in the square root  to be small. Expanding in it yields
\bea
M^2_{u_l}&=&\frac{m^2_2 m_3^2}{M_1^2+m_2^2+m_3^2}\left[1+\mathcal{O}\left(\frac{m^2_2 m_3^2}{\left(M_1^2+m_2^2+m_3^2\right)^2}\right)\right]\\
M^2_{U_h}&=&\left(M_1^2+m_2^2+m_3^2\right)\left[1+\mathcal{O}\left(\frac{m^2_2 m_3^2}{\left(M_1^2+m_2^2+m_3^2\right)^2}\right)\right].
\eea
Matching the lighter mass eigenvalue to the SM quark mass thus requires $m_2m_3/M_1^2\ll1$. Note that the $m_2$ and the $m_3$ arise from mass terms with inherently different symmetry properties. The $m_2$ term links a fundamental 4-plet to a composite $SO(4)$ singlet while the $m_3$ term links a fundamental singlet to a composite 4-plet. Therefore, $m_2$ and $m_3$ are independent parameters which are not required to be of the same order of magnitude by naturalness.
For $m_2,\,M_1\gg m_3$, the left-handed part of the light mass eigenstate becomes a linear combination of the elementary $u_L$ and the composite $\tilde{U}_L$ (\ie left-handed partial compositeness). Such a mixing is strongly constraint because in this case, the coupling of the lighter quark mass eigenstate to the $Z$ boson is modified.\footnote{For partners of the light quarks, this implies a modification of the hadronic width of the $Z$. For $b$ quark partners  $Z\rightarrow b\bar{b}$ is modified at tree level.} We therefore focus on right-handed partial compositeness, where $m_3,\,M_1\gg m_2$.
In what follows, we take $m_3$ as a free $\mathcal{O}(M_1)$ parameter and neglect the effect of $m_2$ apart from fixing it in order to reproduce the correct quark mass in $m_{u_l}$ in the above. In this limit, the heavy quark mass becomes
\beq
M_{U_h}=\sqrt{M^2_1+m^2_3},
\label{pcmassh}
\eeq 
and the bi-unitary transformation which diagonalizes the mass matrix is the identity on the left-handed up-quarks and a rotation by $\varphi$ on the right-handed up-quarks where
\beq
\tan{\varphi}=\frac{m_3}{M_1}.
\eeq

The couplings of the mass eigenstates to the $Z$ bosons follow from rewriting 
\beq
\mathcal{L}_Z =  (\bar{u}_L , \bar{\tilde{U}}_L) 
\left[
\frac{g}{2 c_w}\left(
\begin{array}{cc}
1 & 0 \\
0 & 0
\end{array}
\right)-\frac{2 g}{3}\frac{s^2_w}{c_w}\cdot \mathbbm{1}\right] \slashed{Z}
\left(\begin{array}{c}
u_L\\
\tilde{U}_L
\end{array}
\right) - \frac{2 g}{3}\frac{s^2_w}{c_w}(\bar{u}_R, \bar{\tilde{U}}_R)\slashed{Z}\cdot \mathbbm{1}\left(\begin{array}{c}
u_R\\
\tilde{U}_R
\end{array}
\right), 
\label{ZLag}
\eeq
in the mass eigenbasis  $(u_l,U_h)$. Note that the couplings arising from the $U(1)_X$ gauge couplings are universal, and a rotation into the mass eigenbasis of these terms does not induce any ``mixed'' interactions of the $Z$ to $u_l$ and $U_h$. Such an interaction can only arise from the $SU(2)_L$ component of the $Z$ in the first term, but as the mixing in the left-handed sector is small (of order $m_2/M_1$), the ``mixing'' couplings are negligible. The same holds for mixing couplings of the $W$ gauge boson which follow from the analogous charged current interactions.

The Higgs couplings to the quark mass eigenstates follow from
\beq
\mathcal{L}_{\rm Yuk} =  -\lambda_3 h \bar{\tilde{U}}_L u_R+\mbox{h.c.} = -\lambda^{\rm eff}_{\rm mix}h\bar{U}_{h,L}u_{l,R}- \lambda^{\rm eff}h\bar{U}_{h,L}U_{h,R}+\mbox{h.c.} 
\eeq
with
\bea
\lambda^{\rm eff}_{\rm mix}&=&\lambda_3 \cos(\varphi)=y_R G'(\epsilon)\frac{M_1}{M_{U_h}},\label{lambdapceff}\\
\lambda^{\rm eff}&=&\lambda_3 \sin(\varphi)=y^2_R G(\epsilon)G'(\epsilon)\frac{f}{M_{U_h}}.
\eea

The coupling $\lambda^{\rm eff}$ does not contribute to the production or decay of the partner quarks, and we neglect it for the effective description of the model. Collecting all other terms, the effective partially composite quark model Lagrangian is given by
\beq
\mathcal{L}_{\rm eff}=\mathcal{L}_{\rm SM}+\bar{U}_h\left(i\slashed{\partial}+ e \frac{2}{3}\slashed{A} -g\frac{2}{3}\frac{s^2_w}{c_w}\slashed{Z}+ g_3 \slashed{G}  \right)U_h - M_{U_h}\bar{U}_hU_h
-\left[\lambda^{\rm eff}_{\rm mix}h\bar{U}_{h,L}u_{l,R}+\mbox{h.c.}\right]. \label{Lpceff}
\eeq

The Lagrangian \eqref{Lpceff} and the definition of the effective coupling \eqref{lambdapceff} has been derived for up-type partners. The analogous calculation for down-type partners yields the same Lagrangian with the charge factors $2/3$ being replaced by $-1/3$ as directly follows from the $U(1)_X$ charge assignments. For illustration we embedded $q_L$, $\psi$ and $u_R$ in the $\bold{5}$ of $SO(5)$ in the above, but from the derivation it is apparent, that the effective Lagrangian holds more generally. As long as the quark partner multiplet contains one $SO(4)$ singlet, and in the limit in which all other partner states are decoupled, the only dependence on the chosen representation enters through the functions $\mathcal{F}$ and $\mathcal{G}$ in Eq.~\eqref{Lelmixpc} via the functions $F$ and $G$  in Eq.~\eqref{pcLag1} into the effective coupling $\lambda^{\rm eff}_{\rm mix}$ in Eq.~\eqref{lambdapceff}. As an illustration, Table~\ref{table:pc} shows the corresponding functions and the effective coupling for an embedding of  $\psi$ and $u_R$ in the $\bold{5}$ and $q_L$ in the $\bf 14$. The extension to other embeddings which contain an $SO(4)$ singlet partner is straight forward.

\bigskip

We emphasize that for concrete realizations of partially composite quarks, the parameters $\lambda^{\rm eff}_{\rm mix}$ and $M_{U_h}$ are correlated. In particular, the mass of the quark partner is $M_{U_{h}}=\sqrt{M_1^2+y_R^2f^2 G(\epsilon)}$, such that a light partner mass $M_{U_{h}}$  implies an upper bound on $M_1$ and $y_R$ which in turn implies an upper bound on $\lambda^{\rm eff}_{\rm mix}$ (for fixed $M_{U_h}$ and $f$, and depending on the embedding chosen). In Sec. \ref{sec:phenopc} and \ref{sec:pcresults}, we study the phenomenological implications of the more general effective model defined by Eq. \eqref{Lpceff} in terms of the parameters $\lambda^{\rm eff}_{\rm mix}$ and $M_{U_h}$ and point out the implications for specific quark embeddings.
\begin{table}
\begin{center}
\begin{tabular}{|l|l|l|l|l|l|}
\hline
& $\mathcal{F}(U_{\rm gs})$ & $\mathcal{G}(U_{\rm gs})$ & $F(\frac{h+v}{f})$ & $G(\frac{h+v}{f})$ & $\lambda^{\rm eff}_{\rm mix}$\\
\hline\hline
$q_L\in\bf{5}$ & $U^{i5}_{\rm gs}$ & $U^{i5}_{\rm gs}$ & $-\sin(\frac{h+v}{f})$ & $\cos(\frac{h+v}{f})$ & $\phantom{=}-y_R \sin(\epsilon)\frac{M_1}{M_{U_h}}$\\ 
 $\psi\in\bf{5}$ &&&&&\\
\hline
$q_L\in \bf{14}$ & $U^{i5}_{\rm gs}U^{j5}_{\rm gs}$  & $U^{i5}_{\rm gs}U^{j5}_{\rm gs}$ &  $-\frac{1}{\sqrt{2}}\sin(2 \frac{(h+v)}{f})$ &  $\frac{1}{2\sqrt{5}} ( 5\cos^2( \frac{h+v}{f})-1)$ & $\phantom{=}-\frac{y_R \sqrt{5}}{2} \sin(2\epsilon)\frac{M_1}{M_{U_h}}$\\ 
$\psi\in {\bf 5}$ &&&&& \\ 
\hline
\end{tabular}
\end{center}
\caption{Relations between the effective Lagrangian and parameters given in Eqs.~(\ref{pcLag1},\ref{lambdapceff},\ref{Lpceff}) and explicit embeddings of the quark states into the $\bf{5}$ and $\bf{14}$ of $SO(5)$.}
\label{table:pc}
\end{table}


\subsection{Fully composite quark models}\label{sec:FCtheory}

We repeat the analysis presented in the last section for the fully composite quark model. The fermion Lagrangian in this case is \cite{CHus,DeSimone:2012fs}
\beq
\mathcal{L}=\mathcal{L}_{\rm comp}+\mathcal{L}_{\rm el,mix}
\eeq
with
\bea
\mathcal{L}_{\rm comp}&=&i\ \bar{\psi}(D_\mu+i e_\mu)\gamma^\mu\psi+i\
\bar{u}_R\, \slashed{D} u_R -   M_4\bar{Q}Q -M_1\bar{\tilde{U}}\tilde{U}\nonumber\\
&&
+\left(ic_L\, \bar{Q}_L^i d^i_\mu \gamma^\mu \tilde{U}_L+ i c_R\, \bar{Q}_R^i
d^i_\mu \gamma^\mu \tilde{U}_R+\mbox{h.c.}\right)
+\left(i c_1\, \bar{Q}_R^i d^i_\mu \gamma^\mu u_R+\mbox{h.c.}\right)\,,\nonumber\\
&\underrightarrow{M_4\rightarrow \infty}& i\ \bar{\tilde{U}}\slashed{D}\tilde{U}+i\
\bar{u}_R\, \slashed{D} u_R -M_1\bar{\tilde{U}}\tilde{U},
\label{fcLag1}
\eea
and
\bea
\mathcal{L}_{\rm el,mix}&=&
i\ \bar{q}_L\,\slashed{D}q_L -\left[y_L\, f\bar{q}^5_L \mathcal{F}(U_{\rm gs}) {Q}_R + \mbox{h.c.}\right]\nonumber\\
&& -\left[ y_L\, c_2\, f\bar{q}_L^5 \mathcal{G}(U_{\rm gs}) u_R +  y_L\, c_3\, f\bar{q}_L^5 \mathcal{G}(U_{\rm gs}) \tilde{U}_R  + \mbox{h.c.}\right],\\
&\underrightarrow{M_4\rightarrow \infty}&i\ \bar{q}_L\,\slashed{D}q_L -\left[ \frac{y_L\, c_2\, f}{\sqrt{2}}G(\frac{h+v}{f})\bar{u}_L u_R +\frac{y_L\, c_3\, f}{\sqrt{2}}G(\frac{h+v}{f})\bar{u}_L \tilde{U}_R + \mbox{h.c.}\right]\nonumber,\\
&=&i\ \bar{q}_L\,\slashed{D}q_L-\left[m_2  \bar{u}_L u_R+m_3  \bar{u}_L \tilde{U}_R+\lambda_2 h \bar{u}_L u_R+\lambda_3 h \bar{u}_L \tilde{U}_R+  \mbox{h.c.} +\mathcal{O}(\epsilon^2)\right],\nonumber\\
\label{fcLag2}
\eea
where  (for $q_L$ and $\psi$ in the $\bf 5$)  $\mathcal{F}=U_{\rm gs}^{Ii}$, $\mathcal{G}=U_{\rm gs}^{I5}$, $G=\sin((h+v)/f)$. The expansion of $G$ around $\epsilon=v/f$ yields the effective parameters
\bea
m_{2,3} &=&  \frac{y_L\, c_{2,3}\, f}{\sqrt{2}}G(\epsilon)= -\frac{y_L\, c_{2,3}\, v}{\sqrt{2}}\left[1+\mathcal{O}(\epsilon^2)\right], \\
\lambda_{2,3} &=& \frac{y_L\, c_{2,3}}{\sqrt{2}}G'(\epsilon) =  -\frac{y_L\, c_{2,3}}{\sqrt{2}}\left[1+\mathcal{O}(\epsilon^2)\right] \label{lamfcdef}
\eea
The mass terms read
\beq
\mathcal{L}_m=-(\bar{u}_L, \bar{\tilde{U}}_L) M_u \left(\begin{array}{c} u_R\\ \tilde{U}_R \end{array}\right)+ \mbox{h.c.}
\mbox{\hspace{20pt} with \hspace{20pt}}
M_u= \left(
\begin{array}{cc}
m_2 & m_3\\
0 & M_1
\end{array}
\right).
\eeq
Note that there is no symmetry argument differentiating between the $m_2$ and $m_3$ term, implying that they should be of the same order. The lighter eigenvalue of the mass matrix has to be identified with the SM quark mass. To obtain a small eigenvalue (for first or second family quarks or the bottom), we have to choose $m_{2}\ll M_1$, implying that $m_3$ is also naturally small. Then, the mass eigenvalues are approximately given by
\bea
m_{u_l}= m_2+\mathcal{O}(m_2/M_1)  &\mbox{\hspace{10pt} and \hspace{10pt}}& M_{U_h}\approx \sqrt{M_1^2+m_3^2}\approx M_1.
\eea
The mass matrix is diagonalized by a rotation $T_R=\mathbbm{1}+\mathcal{O}(m^2_3/M^2_1)$ while in the left-handed quark sector, the mixing angle is to leading order given by $\tan(\varphi)=m_3/M_{U_h}$. The situation is therefore opposite to the partially composite case: The right-handed mass eigenstates are gauge eigenstates while the left-handed gauge eigenstates mix. 

Analogous to the partially composite setup, the couplings of the quark mass eigenstates follow from Eq.~\eqref{ZLag} when transforming into the mass eigenbasis $(u_l,U_h)$, which for fully composite quarks yields
\bea
\mathcal{L}_Z&=&\mathcal{L}_{Z,\rm SM}-\frac{2g}{3}\frac{s^2_w}{c_w}\bar{U}_h \slashed{Z} U_h-\frac{g}{2c_w}\sin^2(\varphi)\, \bar{u}_{l,L} \slashed{Z} u_{l,L}+\frac{g}{2c_w}\sin^2(\varphi)\, \bar{U}_{h,L} \slashed{Z} U_{h,L}\nonumber\\
&&\phantom{\mathcal{L}_{Z,\rm SM}}-\left[\frac{g^{\rm eff}_{\rm mix}}{2c_w} \bar{U}_{h,L} \slashed{Z} u_{l,L}+\mbox{h.c.}\right],\label{ZLag2}
\eea
where $\mathcal{L}_{Z,\rm SM}$ contains the SM interaction of the light quark with the $Z$ boson, and 
\beq
g^{\rm eff}_{Z,\rm mix}=g \sin(2\varphi)/2\approx g \frac{m_3}{M_{U_h}}.
\eeq
The third term in Eq.~\eqref{ZLag2} is suppressed by $\mathcal{O}(m^2_q/M_{U_h}^2)$ and hence does not lead to relevant corrections of the partial $Z$ width. The ``mixing'' interaction in the last line of Eq.~\eqref{ZLag2} is small (of order $\mathcal{O}(m_q/M_{U_h})$), but it will play an important role in the determination of the branching ratios of $U_h$ decays.

Analogously, the couplings of the quarks to the $W$ boson are described by
\beq
\mathcal{L}_W=\mathcal{L}_{W,\rm SM}+\left[-\frac{g}{\sqrt{2}} 2 \sin^2(\varphi/2)\bar{d}_{L}\slashed{W}u_{l,L}-\frac{g^{\rm eff}_{W,mix}}{\sqrt{2}}\bar{d}_{L}\slashed{W}U_{h,L} +\mbox{h.c.}\right],
\eeq 
with
\beq
g^{\rm eff}_{W,\rm mix}=g \sin(\varphi) \approx g \frac{m_3}{M_{U_h}}\approx g^{\rm eff}_{Z,\rm mix}.
\eeq
Finally, the ``mixed'' Higgs interactions follow from
\beq
\mathcal{L}_h=-\lambda_3 h \bar{u}_L\tilde{U}_R+\mbox{h.c.}= -\lambda^{\rm eff}_{\rm mix} h \bar{u}_{l,L}U_{h,R}+\mathcal{O}(m_3/M_{U_h})+\mbox{h.c.}\, .
\eeq
with $\lambda^{\rm eff}_{\rm mix}=\lambda_3$. To simplify the discussion, let us note that
\bea
\frac{g^{\rm eff}_{Z,\rm mix}}{2 c_w\lambda^{\rm eff}_{\rm mix}}\approx\frac{g}{2 c_w}\frac{m_3}{M_1}\frac{f G(\epsilon)}{m_3 G'(\epsilon)}
=\frac{g}{2 c_w}\tan(\epsilon)\frac{f}{M_1}\approx \frac{g}{2 c_w}\frac{v}{f}\frac{f}{M_1}=\frac{m_Z}{M_1}.\label{eq:geffapprox}
\eea
Using this relation, the effective Lagrangian for  $SO(4)$ singlet partners of fully composite light quarks reads
\bea
\mathcal{L}_{\rm eff}&=&\mathcal{L}_{\rm SM}+\bar{U}_h\left(i\slashed{\partial}+ e \frac{2}{3}\slashed{A} -g\frac{2}{3}\frac{s^2_w}{c_w}\slashed{Z}+ g_3 \slashed{G}  \right)U_h - M_{U_h}\bar{U}_h U_h \nonumber\\
&&-\left[\lambda^{\rm eff}_{\rm mix}\frac{m_Z}{M_1}\bar{U}_{h,L}\slashed{Z}u_{l,L}+\sqrt{2}\lambda^{\rm eff}_{\rm mix}\frac{m_W}{M_1}\bar{U}_{h,L}\slashed{W}d_{L}+\lambda^{\rm eff}_{\rm mix}h\bar{U}_{h,L}u_{l,R} +\mbox{h.c.}\right]. \label{Lfceff}
\eea
 
The ``mixed'' interactions are suppressed by $m_q/M_1$ and therefore do not play a role for the production of composite quark partners, which is thus dominated by QCD pair production. They are important for the branching fractions of the $U_h$, however. The partial widths for the decays $U_h\rightarrow X j$ are given by 
\bea
\Gamma_{U_h\rightarrow h u_l}&=& M_{U_h} \frac{|\lambda^{\rm eff}_{\rm mix}|^2}{32 \pi}\left(1-\frac{m_h^2}{M_{U_h}^2}\right)^2,\nonumber\\
\Gamma_{U_h\rightarrow Z u_l}&=& \frac{M_{U_h}^3}{M_1^2} \frac{|\lambda^{\rm eff}_{\rm mix}|^2}{32 \pi}\left(1-\frac{m^2_Z}{M_{U_h}^2}\right)^2 \times\left(1+\frac{2 m_Z^2}{M_{U_h}^2}\right),\nonumber\\
\Gamma_{U_h\rightarrow W d}&=&  \frac{M_{U_h}^3}{M_1^2}  \frac{|\lambda^{\rm eff}_{\rm mix}|^2}{16 \pi}\left(1-\frac{m^2_W}{M_{U_h}^2}\right)^2 \times\left(1+\frac{2 m_W^2}{M_{U_h}^2}\right).\label{eq:width}
\eea
Thus, up to corrections of order $\mathcal{O}(m^2_{W/Z/h}/M_{U_h}^2)$ the branching ratios of the $U_h$  for decays into $W$, $Z$, and $h$ plus jet are are $50\%$, $25\%$ and $25\%$.

\bigskip

Like for partially composite models, the Lagrangian \eqref{Lfceff} holds also for down-type partners when replacing the charge factors $2/3$ with $-1/3$. Again, the effective Lagrangian can be applied to other embeddings of $q_L$ and $\psi$ into $SO(5)$, as long as the quark partner contains an $SO(4)$ singlet and as long as the other partner states are decoupled. As an example, we list the respective functions and the expression for the effective coupling for $q_L$ embedded in the $\bf 14$ and $\psi$ embedded in the $\bf 5$ of $SO(5)$ in Table~\ref{table:fc}.

\begin{table}
\begin{center}
\begin{tabular}{|l|l|l|l|}
\hline
                        & $\mathcal{G}(U_{\rm gs})$ &  $G(\frac{h+v}{f})$ & $\lambda^{\rm eff}_{\rm mix}= \lambda_3$\\
\hline\hline
$q_L\in\bf{5}$,  $\psi\in\bf{5}$ & $U^{i5}_{\rm gs}$  & $-\sin(\frac{h+v}{f})$ & $-\frac{y_Lc_3}{\sqrt{2}}\cos(\epsilon)$\\ 
\hline
$q_L\in \bf{14}$, $\psi\in {\bf 5}$ & $U^{i5}_{\rm gs}U^{j5}_{\rm gs}$  &  $-\frac{1}{\sqrt{2}}\sin(2\frac{(h+v)}{f})$ & $-y_Lc_3\cos(2 \epsilon)$  \\ \hline
\end{tabular}
\end{center}
\caption{Relations between the effective Lagrangian and parameters given in Eqs.~(\ref{fcLag2},\ref{lamfcdef}) and explicit embeddings of the quark states into the $\bf{5}$ and $\bf{14}$ of $SO(5)$.}
\label{table:fc}
\end{table}

\newpage

\subsubsection{$U_h$ decays in the presence of higher dimensional operators}\label{sec:FCHDops}

The $U_h$ decay rates determined in Eq.(\ref{eq:width}) are proportional to $|\lambda^{\rm eff}_{\rm mix}|^2$. As opposed to the partially composite setup discussed in Sec. \ref{sec:PCtheory}, the mixing parameters in the fully composite setup are strongly suppressed, $|\lambda^{\rm eff}_{\rm mix}|\propto \mathcal{O}(m_q/v)$, in order to reproduce the quark masses. Therefore, higher dimensional operators in the strongly coupled sector which are induced by integrating out other strongly coupled resonances can play the dominant role for $U_h$ decays and strongly modify the branching ratios.\footnote{For a more extensive discussion of bounds on strongly coupled vector resonances in models with right-handed quark compositeness \cf Ref.~\cite{Redi:2013eaa}.}

In the absence of additional symmetries, the lightest colored vector resonance $\rho$ can couple to $U_h$ and $u_R$ via
\beq
\mathcal{L}_{comp}\supset i g_{\tilde{U}\tilde{U}}\bar{\tilde{U}}\slashed{\rho} \tilde{U}+ i g_{\tilde{U}u}\bar{\tilde{U}}_R\slashed{\rho} u_R+ i g_{uu}\bar{u}_R\slashed{\rho} u_R + \mbox{ h.c. },
\eeq
which, upon integrating out $\rho$, yields a four-fermion operator
\beq
\mathcal{L}_{comp}\supset \lambda_{Uuuu} ( \bar{\tilde{U}}_R\gamma^\mu u_R ) \left( \bar{u}_R\gamma_\mu u_R \right),
\eeq 
implying a possible decay $U_h\rightarrow uuu$.\footnote{Integrating out massive colored scalar or pseudo-scalar resonances leads to analogous four-fermion interactions.} The resulting partial decay width for $U_h\rightarrow uuu$ is 
\beq
\Gamma_{U_h\rightarrow uuu}\sim M_{1} \frac{(g_{\tilde{U}u}g_{uu})^2}{3456 \, \pi^3} \left(\frac{M_1}{M_\rho}\right)^4
\eeq
which exceeds the partial decay widths Eq.(\ref{eq:width}) for all light quark flavors. Therefore in typical fully composite light quark models, $U_h$ decays into three jets, making a Higgs final state search obsolete.
The three-jet decay, as well as the $d$-term interactions in Eq.(\ref{fcLag1}), violate a $U(1)$ symmetry $u_R\rightarrow e^{i\phi}u_R$ which is present at the leading order composite Lagrangian. If we make a very special UV-dependent assumption that the strong dynamics preserves this symmetry, enforcing $g_{\tilde{U}u}=0$, then the Yukawa terms in $\mathcal{L}_{mix}$ Eq.(\ref{fcLag2}) are the only source of chiral symmetry breaking. In this case, in the limit $y_L\rightarrow 0$, the $U(1)$ symmetry is restored and forbids the decay of $U_h$ into SM particles. As a result of it, any operator that induces $U_h$ decay must be suppressed by $y_L$, on top of additional suppression factors for higher dimensional or loop-induced operators. 

Under the special assumption that the strongly coupled sector does not break the $U(1)$ symmetry, the leading operators are therefore those given in Eq.~(34), and the $U_h$ decays into $Wu$, $Zu$, and $hu$ with branching ratios of $\sim 50\%$, $\sim 25\%$, and $\sim 25\%$ while for a generic strongly coupled sector, the main decay channel is $U_h\rightarrow uuu$.


\newpage

\section{Phenomenology of composite quarks with $SO(4)$ singlet partners}\label{sec:pheno}

The phenomenology of quark partners in partially and fully composite quark models, which is due to the different allowed parameter ranges of  $\lambda^{\rm eff}_{\rm mix}$, as well as the differing branching ratios for partner states decaying into $h j$, $W j$ and $Z j$ final states. In the following, we discuss the production, decay, and different search channels for both setups.


\subsection{Quark partners in partially composite models}\label{sec:phenopc}

Partners states of partially composite light quarks couple to gluons via the strong interaction, and to the Higgs and light quarks via the coupling $\lambda^{\rm eff}_{\rm mix}$. This mixing coupling can be sizable whilst still reproducing a small mass for the SM quark. 

\begin{figure}
\begin{center}
\includegraphics[scale=.93]{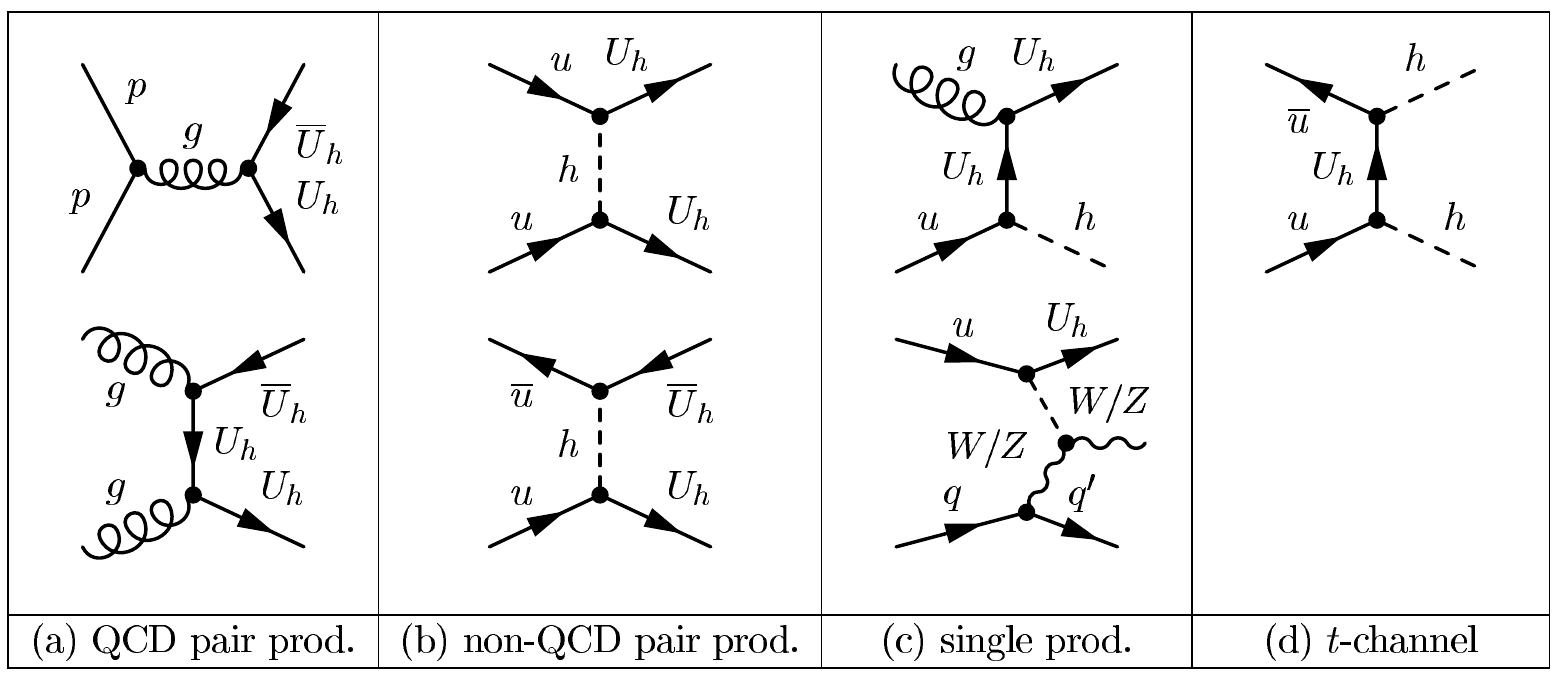} 
\end{center}
\caption{Main production channels of partners of partially composite up-quarks and Higgs bosons. The production channels for $d,s,c,$ and $b$ partners are analogous.}
\label{fig:prod}
\end{figure}

The main production channels involving the composite light quark partner states are shown in Fig.~\ref{fig:prod}.\footnote{Several of these production channels have been discussed in Refs.~\cite{Vignaroli:2012nf,Vignaroli:2012sf} as well as in Ref.~\cite{Atre:2013ap} in the context of the degenerate bidoublet model.} We present the graphs involving an up-type partner. The graphs for other light quark partners are analogous. The QCD pair production channels shown in panel~(a) are independent of quark flavor and of the value of the effective coupling $\lambda^{\rm eff}_{\rm mix}$. Their cross section only depends on the partners' mass $M_{U_h}$. The non-QCD pair production channels shown in panel (b) on the other hand depend on both, quark flavor (\ie the parton distribution functions (PDFs) of the initial $qq$, $q\bar{q}$, or the not shown $\bar{q}\bar{q}$ quark pairs) as well as on the size of the respective $\lambda^{\rm eff}_{\rm mix}$. The same holds true for the single production channels shown in panel (c), where for a light quark partners, the upper diagram yields the dominant contribution. Finally, quark partners modify the di-Higgs production cross section via the $t$-channel process shown in Fig.~\ref{fig:prod} (d). Again, this process depends on both, the partner quark species, as well as on $\lambda^{\rm eff}_{\rm mix}$.

The only available decay channel for $SO(4)$ singlet partners in partially composite models is a decay into a Higgs and a SM quark. For partners of the $u,d,s$, and $c$ quark, this implies a Higgs and a jet per produced quark partner in the final state,\footnote{Note that jet-flavor-tagging targeted towards charm-flavored jets may also improve the charm partner searches in the future. (\cf \eg ~\cite{Delaunay:2013pja}).} while for partners of the third family, the final state top or bottom can be tagged.

\begin{figure}
\begin{center}
\begin{tabular}{cc}
\includegraphics[scale=0.8]{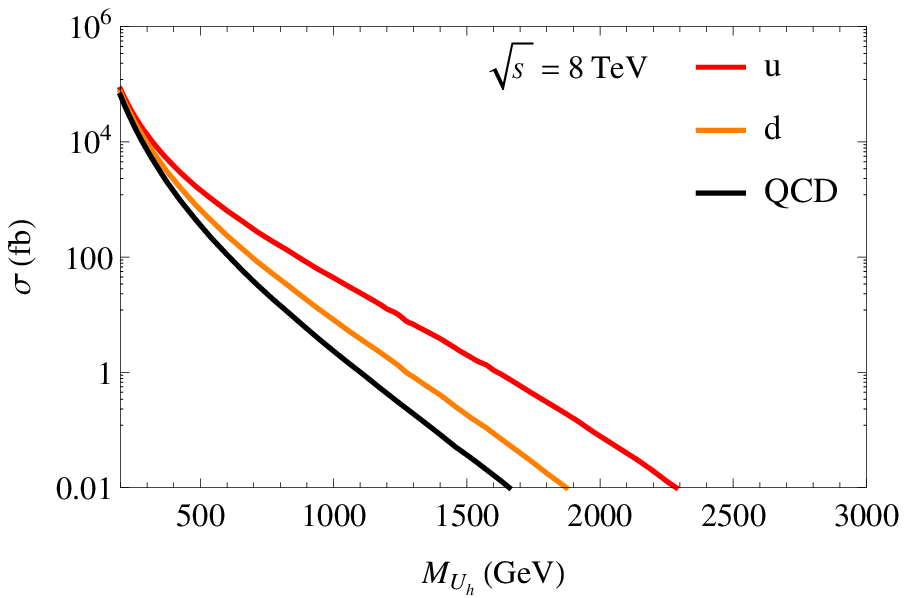} &
\includegraphics[scale=0.8]{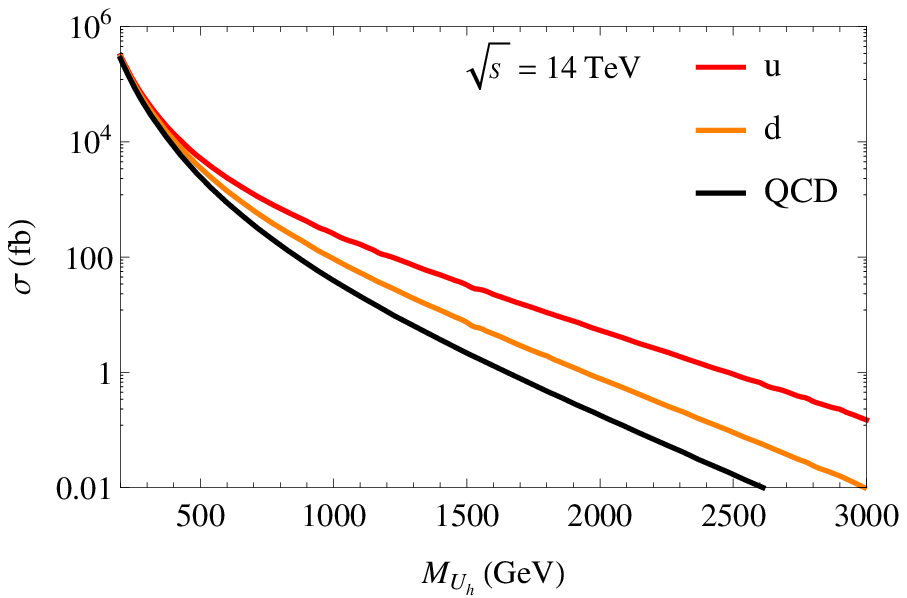}
\end{tabular}
\end{center}
\caption{Production cross section for a pair of quark partners in partially composite quark models as a function of the partners' mass $M_{U_h}$ for LHC at $8 \tev$ (left) and $14 \tev$ (right). 
The first two lines from the top correspond to the pair production cross section with $\lambda^{\rm eff}_{\rm mix}=1$ for partners of the up (red) and down (orange),
while the third line (black) denotes the QCD pair production cross section. The non-QCD pair production cross sections with $\lambda^{\rm eff}_{\rm mix}=1$ for partners of the $s,c$ and $b$ quarks are PDF suppressed. Thus, the pair production cross section for these quark partners is to a good approximation given by the QCD pair production cross section.
}
\label{fig:prodXsec}
\end{figure}

A very promising  signature for the discovery of $SO(4)$ singlet partners is therefore two Higgses with two associated high $p_T$ jets or bottoms.  Fig.~\ref{fig:prodXsec}  shows the cross section for this process as a function of the partner mass $M_{U_h}$.  Here, we assume the presence of a partner of only one SM quark. The QCD production channel provides a $\lambda^{\rm eff}_{\rm mix}$-independent contribution to the cross-section, shown as the black line (the lowest of the three curves). The upper two lines indicate the QCD plus non-QCD cross-section, assuming a reference value of $\lambda^{\rm eff}_{\rm mix}=1$ for the respective quark partner. For $u$ (first line from the top, red) and $d$ (second line from the top, orange) quark partners, the non-QCD contribution can be substantially increased, while for other quark partners, the non-QCD contribution is PDF suppressed. Bearing in mind that the non-QCD cross section scales like $\sim(\lambda^{\rm eff}_{\rm mix})^4$, the cross sections for other values of $\lambda^{\rm eff}_{\rm mix}$ can easily be inferred.

\begin{figure}
\begin{center}
\begin{tabular}{cc}
\includegraphics[scale=0.8]{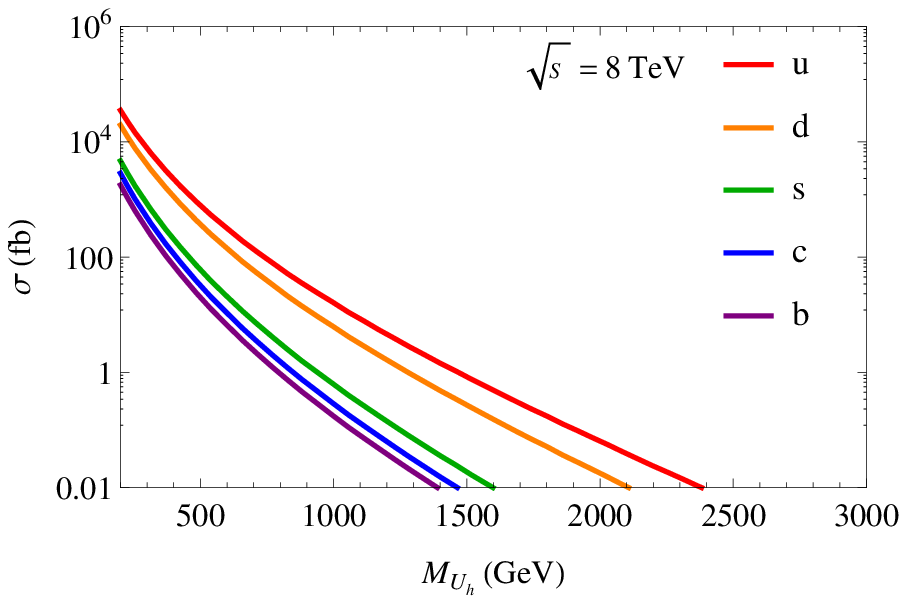} &
\includegraphics[scale=0.8]{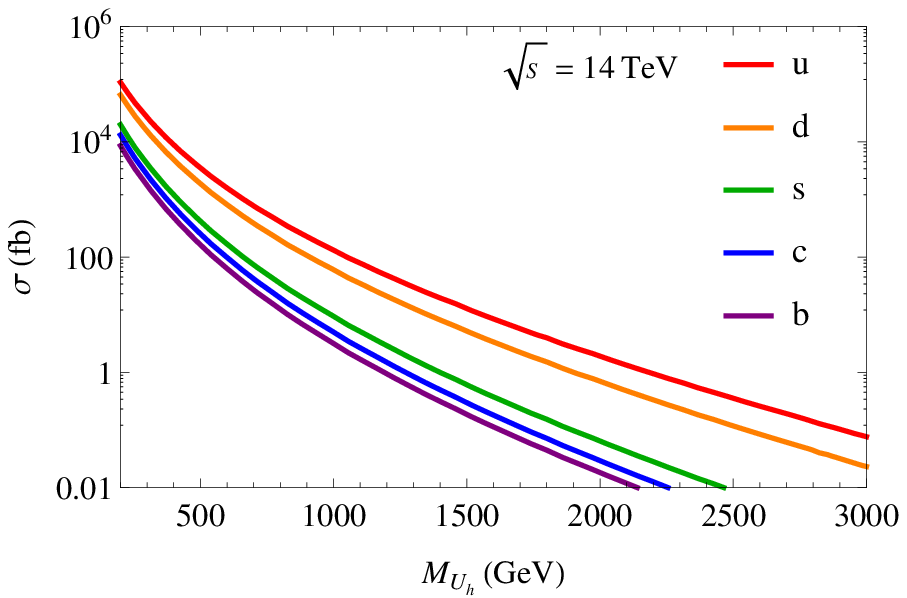}
\end{tabular}
\end{center}
\caption{Production cross section for a single quark partner in partially composite quark models as a function of the partners' mass $M_{U_h}$ for LHC at $8 \tev$ (left) and $14 \tev$ (right).  Lines denote (from right to left): Single production cross section with $\lambda^{\rm eff}_{\rm mix}=1$ for partners of the $u,d,s,c,b$ quark.}
\label{fig:prodXsecsing}
\end{figure}

Singly produced quark partners typically yield a final state with two Higgses and one high $p_T$ jet or $b$-jet. Fig.~\ref{fig:prodXsecsing} shows the cross sections for this channel for a $u,d,s,c$ and $b$ partner with $\lambda^{\rm eff}_{\rm mix}=1$. The single production cross section scales like $\sim(\lambda^{\rm eff}_{\rm mix})^2$.

\begin{figure}
\begin{center}
\begin{tabular}{cc}
\includegraphics[scale=0.8]{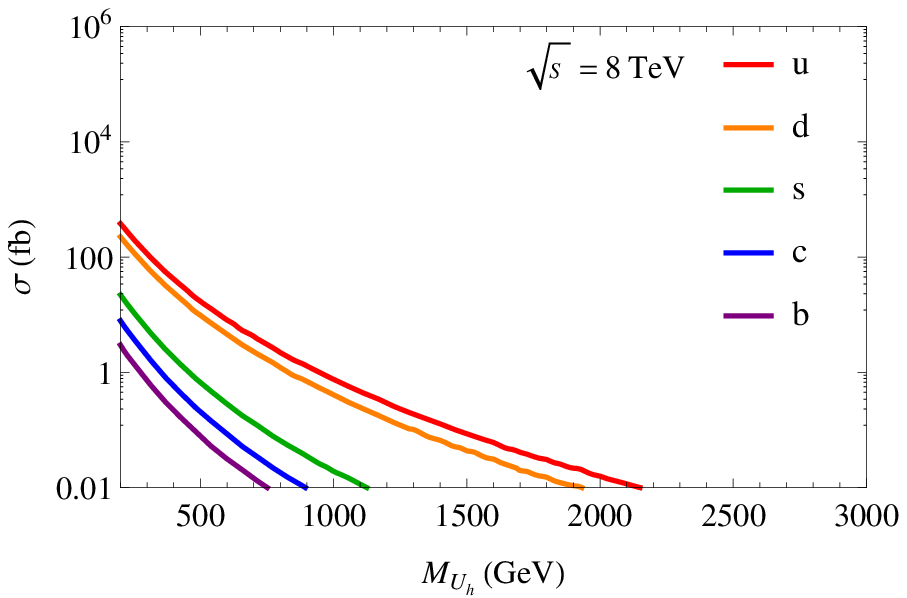} &
\includegraphics[scale=0.8]{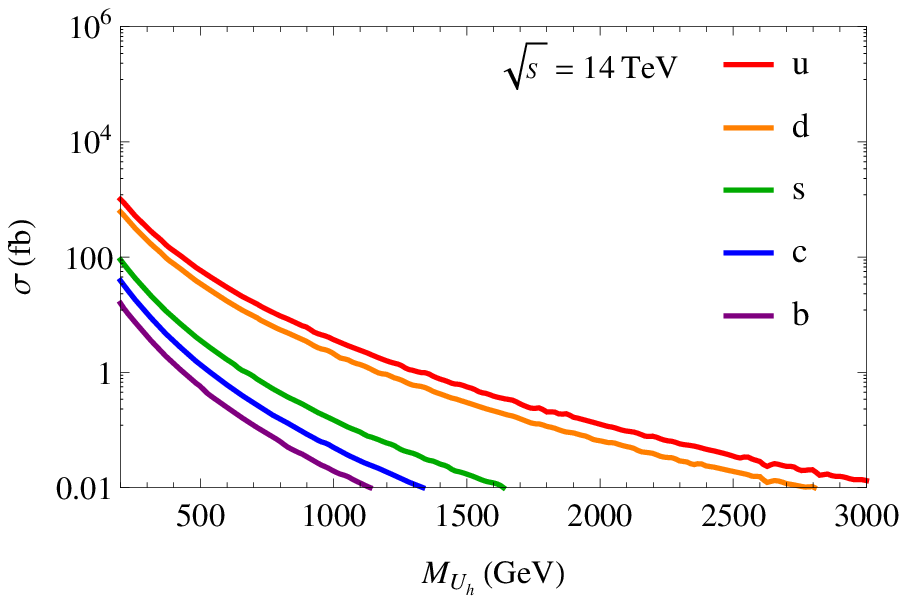}
\end{tabular}
\end{center}
\caption{Production cross section for two Higgses via $t$-channel exchange of a composite quark partner for LHC at $8 \tev$ (left) and $14 \tev$ (right).  Lines denote (from right to left): Cross section with $\lambda^{\rm eff}_{\rm mix}=1$ for the exchange of $u,d,s,c,b$ quark partners.}
\label{fig:prodXseczero}
\end{figure}

Finally, the $t$-channel exchange of $U_h$ shown in Fig.~\ref{fig:prod} (d) contributes to di-Higgs production, in this case without associated high $p_T$ jets or $b$-jet. The cross section depends on the partner quark species and scales like  $\sim(\lambda^{\rm eff}_{\rm mix})^4$.  Fig.~\ref{fig:prodXseczero} shows the di-Higgs production cross sections associated to this channel for a reference value $\lambda^{\rm eff}_{\rm mix}=1$.
 
\bigskip

From the above discussion it is apparent that the di-Higgs channel (with two, one or zero associated high $p_T$ jets or $b$-quarks) provides a golden search channel for $SO(4)$ singlet partner quarks. However as of now, no direct constraints on the  di-Higgs channel are published by ATLAS or CMS, such that in this article, we attempt to obtain constraints on the model from SM Higgs searches in Sec.~\ref{sec:simulation}. It is also important to notice that the currently available LHC studies for the single Higgs plus jets relevant for our analysis in the next section has been performed for the SM Higgs searches. In the near future, dedicated searches for boosted Higgs signals~\cite{BOOSTEDH} in the currently accumulated LHC data set can significantly improve the bound we determine, here.


\subsection{Partners of fully composite quarks}\label{sec:phenofc}

As shown in Sec.~\ref{sec:FCtheory}, fully composite light quark models naturally require a small coupling $\lambda^{\rm eff}_{\rm mix}$ of order $\sim m_q/v$. Therefore  for quark partners, all non-QCD production processes shown in Fig.~\ref{fig:prod} are suppressed, and the QCD pair production processes Fig.~\ref{fig:prod} (a) are the only production channels relevant for LHC phenomenology.\footnote{For partners of fully composite quarks, additional non-QCD production channels exist because the partners also have effective mixing couplings with electroweak gauge bosons and a light quark. The mixing couplings are however proportional to $\lambda^{\rm eff}_{\rm mix}$  (\cf  Eq.~\eqref{eq:geffapprox}), such that these production channels are also suppressed as compared to QCD production.} Partners of fully composite quarks are therefore produced in pairs. The production cross section is independent of the quark species and insensitive to $\lambda^{\rm eff}_{\rm mix}$. 

In generic fully composite models, a $SO(4)$ singlet partner quark dominantly decays into three jets, as outlined in Sec.~\ref{sec:FCHDops}. In this case, the CMS search for three-jet resonances \cite{Chatrchyan:2013gia} implies a bound of $M_{U_h} \gsim 650$ GeV.\footnote{In \cite{Chatrchyan:2013gia} the bound of $M_{res} > 650$ GeV was obtained for pair produced gluinos which decay into three jets in an $R$-parity violating SUSY scenario, but given the analogous production channels for the fully composite model, a similar bound is to be expected,  if the three-jet decay channel dominates.} 

This bound is avoided when assuming that the strongly coupled sector respects a $U(1)$ symmetry $u_R\rightarrow e^{i\phi} u_R$ which is only broken via the Yukawa terms in Eq.~(\ref{fcLag2}). In this special case, quark partners have dominant decay channels into a SM quarks and a $W$, $Z$, or Higgs boson with the branching ratios of these three channels being $\sim 50\%$, $\sim 25\%$, and $\sim 25\%$, as shown in Sec.~\ref{sec:FCtheory}.
The final states arising from pair production of partners of the first two families (and their approximate branching fractions)  are then $hhjj$ ($1/16$), $hWjj$ ($1/4$), $hZjj$ ($1/8$), $WWjj$ ($1/4$), $WZjj$ ($1/4$), and $ZZjj$ ($1/16$). $b$ partners yield the analogous final states with $b$-jets instead of jets.

Before focussing  in the next section on final states containing at least one Higgs boson, let us  summarize the current status of searches for the final states involving $W$ or $Z$ bosons, and their implications for the $SO(4)$ singlet quark partners in  fully composite quark models. Searches relevant for partners of the first two families include the ATLAS search for the $WWjj$ final state \cite{ATLAS:2012qe}, a recast of the $WWjj$ final state performed in Ref.~\cite{CHus} based on the CMS leptoquark search \cite{cms8lepto}, and the ATLAS inclusive search for two 3-jet resonances \cite{Chatrchyan:2013gia}. A naive comparison of these bounds with the QCD production cross section shown in Fig.~\ref{fig:prod} (times the respective branching ratios given above) indicates no bounds on quark partners of the first and second family quarks. 

Searches relevant for bottom quark partners include the  ATLAS search for two 3-jet resonances with at least one $b$-jet \cite{Chatrchyan:2013gia}, the ATLAS search for pair-produced $b$ quark partners decaying into $Zb$ \cite{ATLAS:13-056}, and the CMS searches  for pair-produced $b$ partners in the lepton+jets final state \cite{CMS-12-019} and in the dilepton+jets final state \cite{CMS-12-021}. The strongest bounds on partners of fully composite $b$ quarks are $M_{U_h} >645 \gev$ \cite{ATLAS:13-056} and $M_{U_h} >700 \gev$ \cite{CMS-12-019}.


\section{Constraining composite light quark partners with Higgs searches}\label{sec:results}

For models with only $SO(4)$ singlet partners of light quarks, the interactions of the light quarks to the Higgs are SM-like.\footnote{For a more detailed discussion of Higgs couplings for the composite Higgs models, see the recent studies~\cite{Azatov:2011qy, Espinosa:2012ir,Carmi:2012in, Moreau:2012da, Delaunay:2013iia, Montull:2013mla}.} Also, the composite quark partner $U_h$ does not yield sizable BSM contributions to the one-loop induced $h\gamma\gamma$ and $hgg$ vertices. The dominant contribution arises from a triangle diagram with $U_h$ in the loop. The coupling $\lambda^{\mathrm{\rm eff}}_{\mathrm{\rm mix}}$ of the Higgs to two $U_h$ quarks is Yukawa-suppressed, such that this contribution is negligible for light quark partners. ``Mixed'' triangle-diagrams with a $U_h$ and a $u_l$ or $d$ in the loop do not exist. As apparent from the Lagrangians Eq.~\eqref{Lpceff} (for partially composite quarks) and Eq.~\eqref{Lfceff} (for fully composite quarks), ``mixing'' gluon or photon interactions with one light and one heavy quark are absent. For the same reasons, one-loop induced BSM corrects to the $hWW$ and $hZZ$ vertices are small as well.\footnote{For partners of fully composite quarks ``mixed'' triangle loops with  a $U_h$ and a $u_l$ or $d$ in the loop do exist, but the diagram is Yukawa suppressed.} Therefore, the couplings of all SM states to the Higgs are insensitive to the presence of $SO(4)$ singlet partners of light quarks. The production channels of the Higgs via gluon fusion and vector boson fusion are SM-like, and that all decay rates of the Higgs are SM-like as long as the quark partner $U_h$ is heavier than the Higgs. Hence for composite light quark partner models, Higgs events of at the LHC can be separated into a SM background and a BSM production of Higgses via the production and decay of heavy quark partners or $t$-channel exchange of heavy quark partners in the processes discussed in Sec.~\ref{sec:pheno}.
 
The events discussed in Sec.~\ref{sec:pheno} increase the Higgs production cross section. Moreover, the topology and kinematical distributions of  events with a Higgses which result from the decay of a $U_h$ quark differ from the SM Higgs events. In the SM processes, the Higgs boson is typically produced at threshold, implying low Higgs $p_T$, while Higgses from $U_h$ decays are boosted when the $U_h$ is heavier than the Higgs. Furthermore, $U_h$ pair production and subsequent decay into $hhjj$ leads to a higher number of jets -- in particular with high $p_T$ -- as compared to the SM processes for Higgs production. Thus, even in the current absence of dedicated searches for di-Higgs events or searches for BSM signals in the invariant mass distribution of the Higgs and the leading jet, measurements of the differential cross section of Higgs events can be used to constrain Higgs events which result from $U_h$ decays.

In Ref.~\cite{ATLAShiggs}, the ATLAS collaboration presented results on differential cross sections of the Higgs in the $h\rightarrow\gamma\gamma$ channel. In particular, Ref.~\cite{ATLAShiggs} studies the $p^{\gamma\gamma}_T$, $N_{\mathrm{jets}}$, and the highest $p^{\mathrm{jet}}_T$ distributions which are in good agreement with the SM predictions. We use the experimental results of Ref.~\cite{ATLAShiggs} in order to derive constraints on the parameter space of partially and fully composite light quark models.


\subsection{Simulation and data evaluation}\label{sec:simulation}

To simulate the BSM contribution to the differential cross sections of the Higgs, we implemented the Lagrangians Eq.~\eqref{Lpceff} (for partially composite quarks) and Eq.~\eqref{Lfceff} (for fully composite quarks) using \textsc{FeynRules 2.0} \cite{Alloul:2013bka,Degrande:2011ua}. For the SM part of the Lagrangian we used the SM implementation provided by \cite{SMImplementation}, interfaced with with the effective Higgs implementation by \cite{HiggsEffectiveImplementation} which we adapted in order to reproduce the total width and the branching ratios of the Higgs for a mass $m_H=125$ GeV given in \cite{Heinemeyer:2013tqa}. With the implementation of the models, we generated parton level Monte Carlo (MC) event samples for the BSM  Higgs production channels discussed in Sec.~\ref{sec:pheno} for proton-proton collision at a center-of-mass energy of 8 TeV, using \textsc{MadGraph 5} \cite{Alwall:2011uj}, interfaced with CTEQ6L1 PDFs \cite{Pumplin:2002vw}. After performing the parton showering and the hadronization with \textsc{Pythia 6.4} \cite{Sjostrand:2006za}, the generator-level MC events have been processed with \textsc{Delphes 3} \cite{deFavereau:2013fsa} and the jet clustering procedure is performed via \textsc{FastJET} \cite{FJ1,FJ2}.

The MC generated data is selected according to the particle level fiducial definitions in Ref.~\cite{ATLAShiggs}. The selection criteria are as follows: the two highest-$E_T$, isolated final state photons, within $|\eta| < 2.37$ and with $105 \gev < m_{\gamma\gamma} < 160 \gev$ are selected. The isolation criterion is the sum of the $p_T$ of all stable particles excluding muons and neutrinos is required to be less than 14 GeV within $\Delta R = \sqrt{(\Delta\eta)^2 +(\Delta\phi)^2} <  0.4$ of the photon. After the pair is selected, a cut on $E_T/m_{\gamma\gamma} > 0.35 (0.25)$ for the two photons is applied. Jets are selected using the anti-$k_t$ jet clustering algorithm \cite{Cacciari:2008gp} with a  distance parameter of $R=0.4$.  The resulting jets are required the have a transverse momentum $p_T  >  30  \gev$, and rapidity $|y|< 4.4$.

\begin{figure}
\begin{center}
\begin{tabular}{cc}
\includegraphics[scale=0.58]{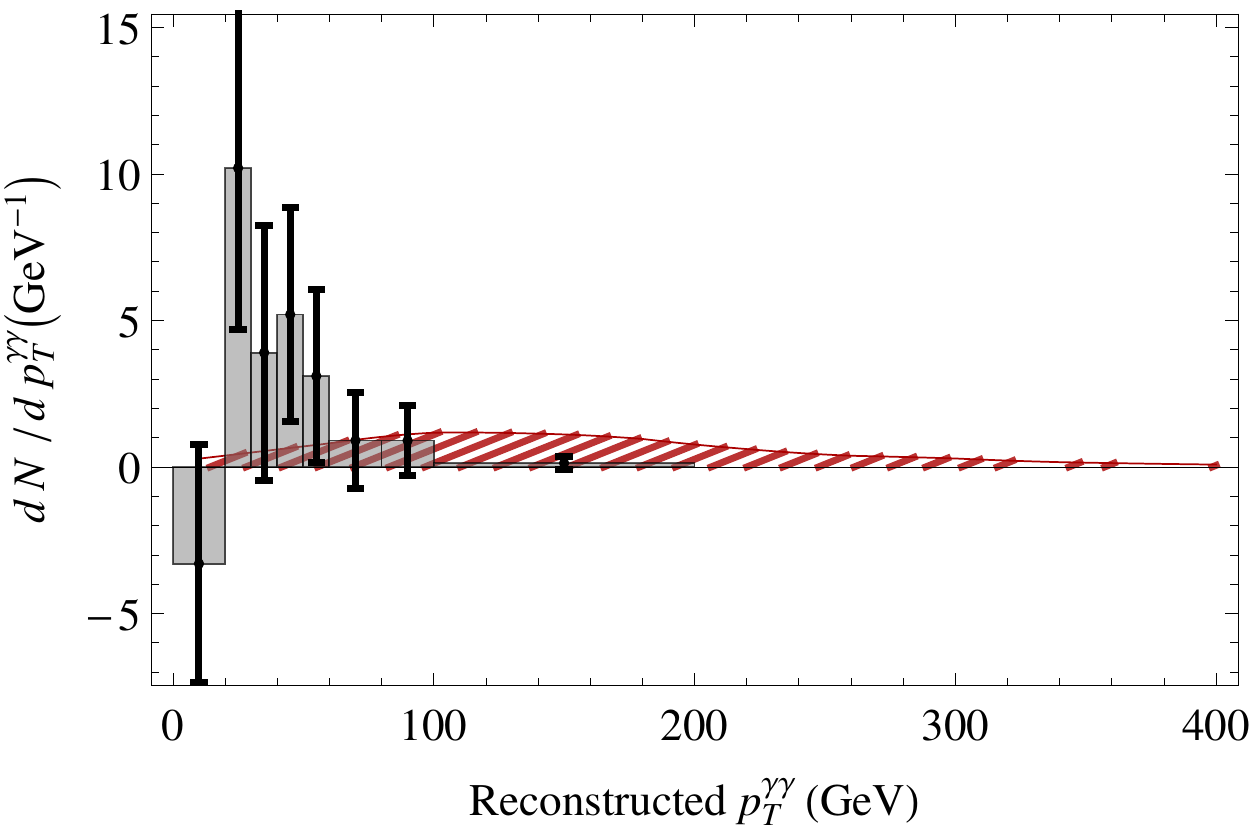} &
\includegraphics[scale=0.58]{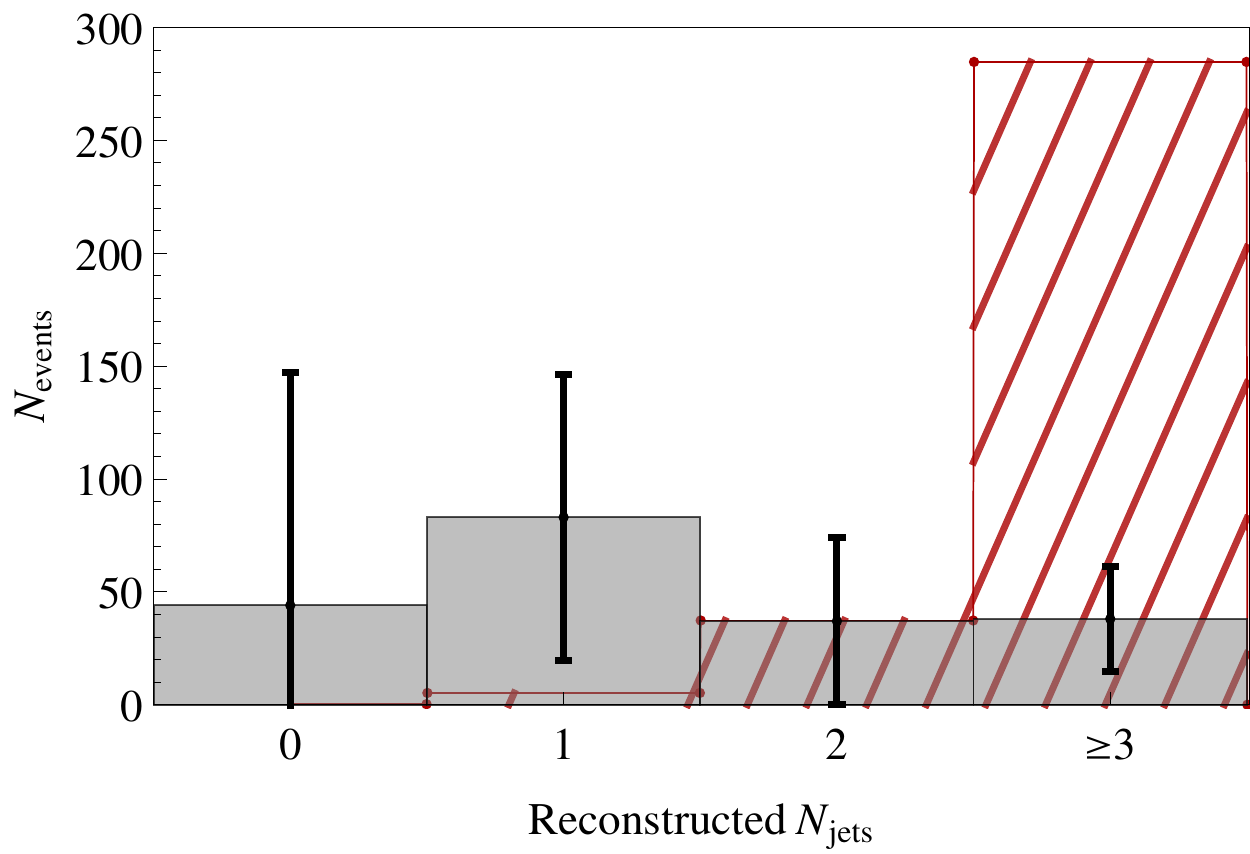}
\end{tabular}
\end{center}
\begin{center}
\begin{tabular}{c}
\includegraphics[scale=0.58]{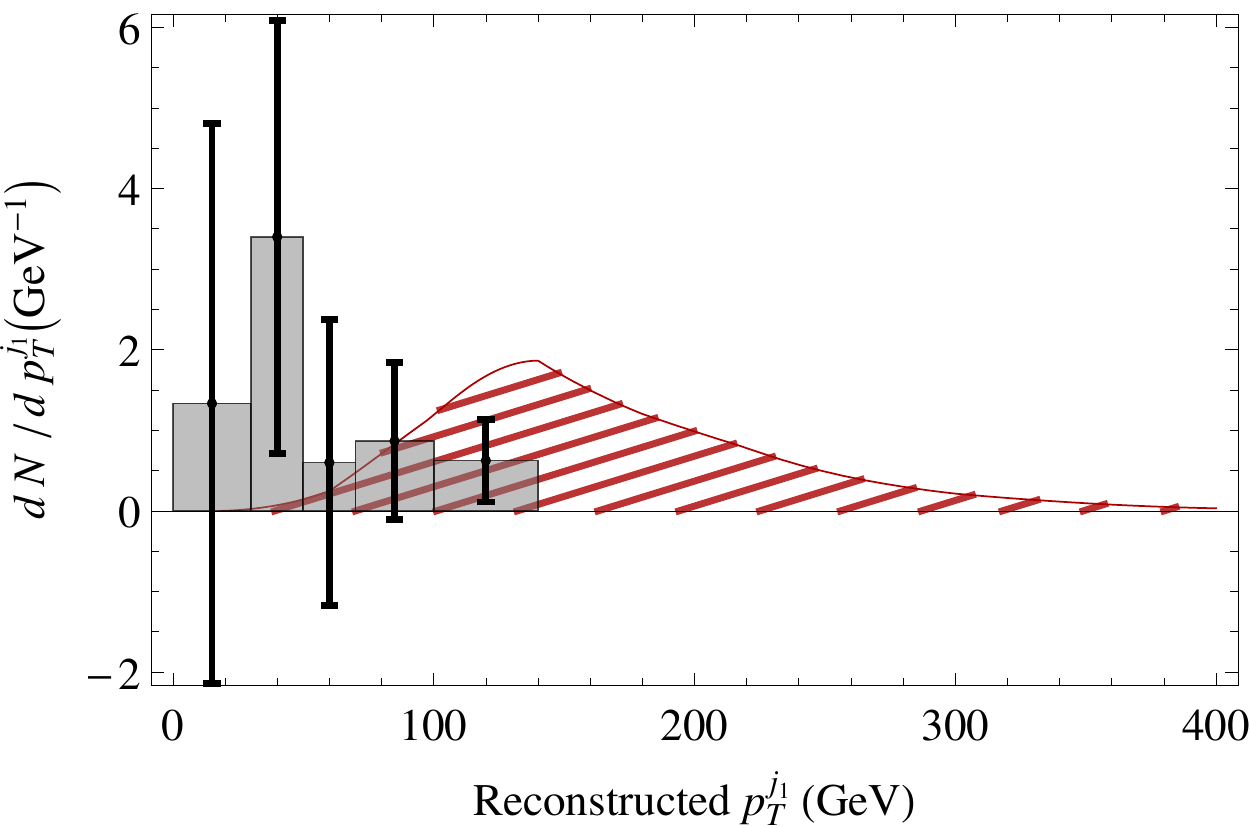} 
\end{tabular}
\end{center}
\caption{Example for the $p_T^{\gamma\gamma}$ (upper left), the $N_{\mathrm{jets}}$ distribution (upper right) and the $p_T^{j_1}$ distribution (center) resulting from a composite quark model.
The distributions shown result from a partially composite down-quark model with a partner mass of $M_{U_h}=300$ GeV and effective coupling of $\lambda^{\rm eff}_{\rm mix} = 1$ (red striped region). The difference between expected SM background as well as the measured yields (gray bins) are taken from Ref.~\cite{ATLAShiggs}.}
\label{fig:exdist}
\end{figure}

With the events which pass the selection criteria outlined above, we simulate the $p_T^{\gamma\gamma}$ and the $N_{\mathrm{jets}}$ distribution as well as the $p_T^{j_1}$ distribution of the most energetic jet of the BSM events.\footnote{The other observables studied in Ref.~\cite{ATLAShiggs} are rapidity $|y^{\gamma\gamma}|$ of the Higgs boson, the helicity angle $|\cos \theta^*|$ in the Collins-Soper frame, the jet veto fractions  $\sigma_{N_{\mathrm{jets}}=i}/\sigma_{N_{\mathrm{jets}}\geq i}$ for jet multiplicities $i$, the azimuthal angle between the leading and the subleading jet $\Delta\phi_{jj}$ and the transverse component of the vector sum of the momenta of the Higgs boson and dijet system $p_T^{\gamma\gamma jj}$. We simulated the corresponding distributions for our BSM channels and found that the effect of composite light quark partners on them is less relevant.}  For $p_T^{\gamma\gamma}$ and $p_T^{j_1}$, the distribution is determined in $p_T$ bins chosen according to the bins in Ref.~\cite{ATLAShiggs}, while for $N_{\mathrm{jets}}$, we consider bins with events with $N_{\mathrm{jets}}=0,1,2$ and $ \geq 3$. The event numbers obtained in each bin at the particle level are divided by bin-by-bin unfolding correction factors 
\begin{equation}
c_i = \frac{n_i^{\mathrm{Particle\;level}}}{n_i^{\mathrm{Reconstructed}}}
\end{equation}
provided in  Ref.~\cite{ATLAShiggs} in order to correct the particle level data to a reconstructed data set. For the following analysis, we assume unfolding correction factors for the experimental data and MC data of the SM and BSM to be equal. As an example, Fig.~\ref{fig:exdist} shows the $p_T^{\gamma\gamma}$, the $N_{\mathrm{jets}}$ and the $p_T^{j_1}$ distribution resulting from a partially composite down-quark partner model with mass $M_{U_h}=300$ GeV and effective coupling of $\lambda^{\rm eff}_{\rm mix} = 1$.

\bigskip

As to be expected, the highest excess over signal occurs in the $p_T^{\gamma\gamma}$ overflow bin. However, Ref.~\cite{ATLAShiggs} used the overflow bin as a control bin and does not provide a unfolding correction factor for it, such that we conservatively ignore the $p_T^{\gamma\gamma}$ overflow bin in our parameter space constraints presented in the remainder of this section. For reference, we also show the constraints resulting from including the $p_T^{\gamma\gamma}$ overflow bin in Appendix~\ref{app:overflow}. Ignoring overflow bins, the dominant excesses arise from the highest $p_T^{\gamma\gamma}$ bin and the $N_{\mathrm{jets}}\geq3$ bin. The highest $p_T^{j_1}$ bin also shows an excess, but it plays less of a role in the determination of constraints on the composite Higgs models, because the error on this bin is large.\footnote{The above statements apply to first and second generation quark partners as well as to bottom quark partners as no bottom tagging is used on the final state jets. Using bottom tagging could strengthen the bottom partner bound considerably as the highest $p_T$ jet in these events is typically a bottom resulting from a partner state decay.}   To obtain a bound on the mass and the coupling of the composite quark models, we perform a simplistic $\chi^2$ test on the $p_T^{\gamma\gamma}$, $p_T^{j_1}$, and  $N_{\mathrm{jets}}$ bins 
\begin{equation}
\chi^2 = \sum_{i}  \frac{\left(\hat{N}_i - N_i\right)^2}{\sigma^2_i},
\end{equation}
where $\hat{N}_i$ is the number BSM events obtained from the MC plus the number of SM events from Ref.~\cite{ATLAShiggs} in the respective bin, and $N_i$ and $\sigma_i$ are the respective measured values and errors.


\subsection{Results for partners of partially composite light quarks}\label{sec:pcresults}

\begin{figure}[h]
\begin{center}
\begin{tabular}{cc}
\includegraphics[scale=0.75]{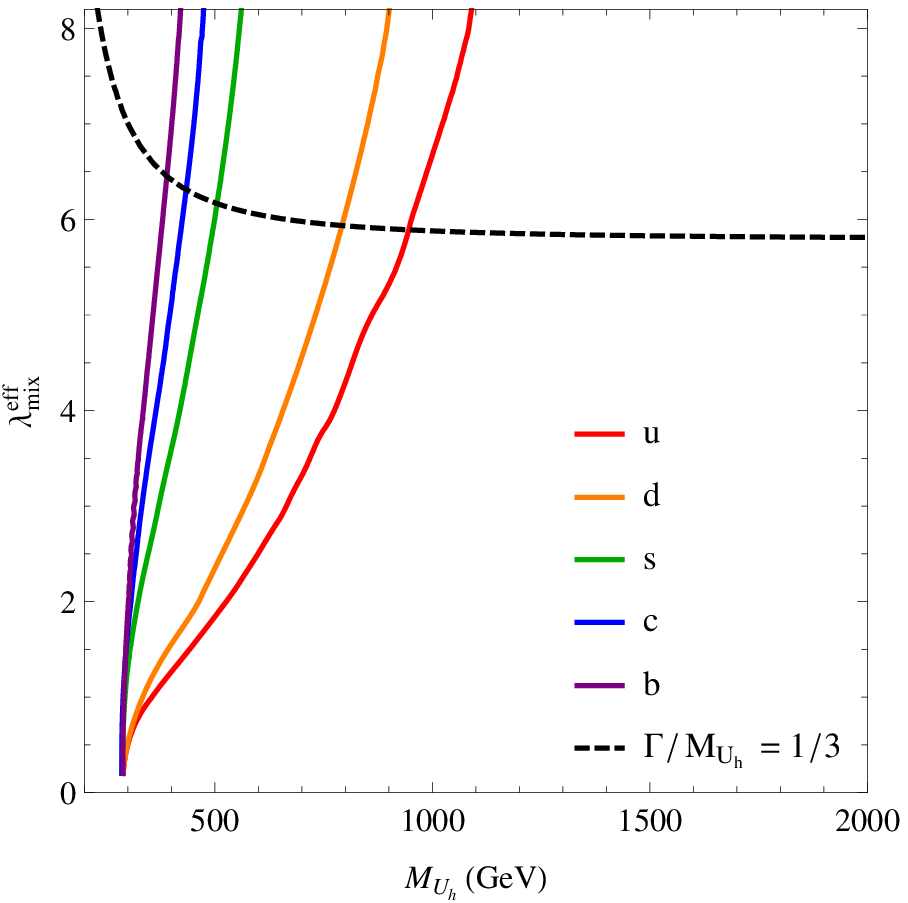} &
\includegraphics[scale=0.75]{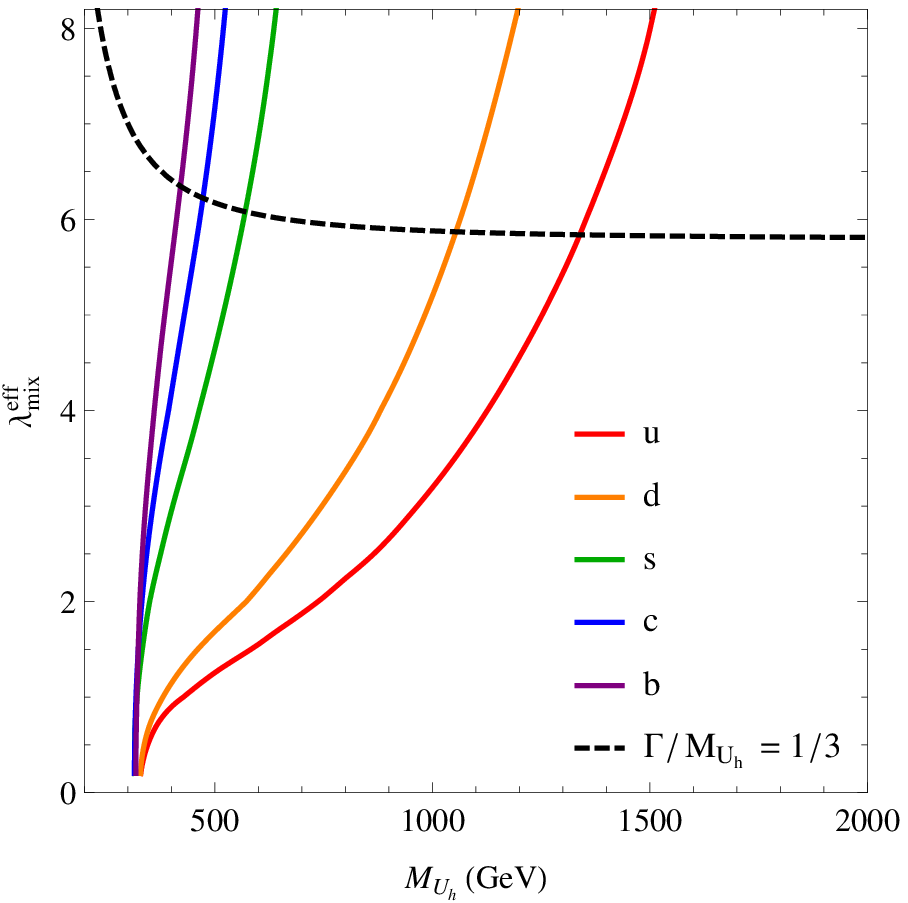}
\\
\includegraphics[scale=0.75]{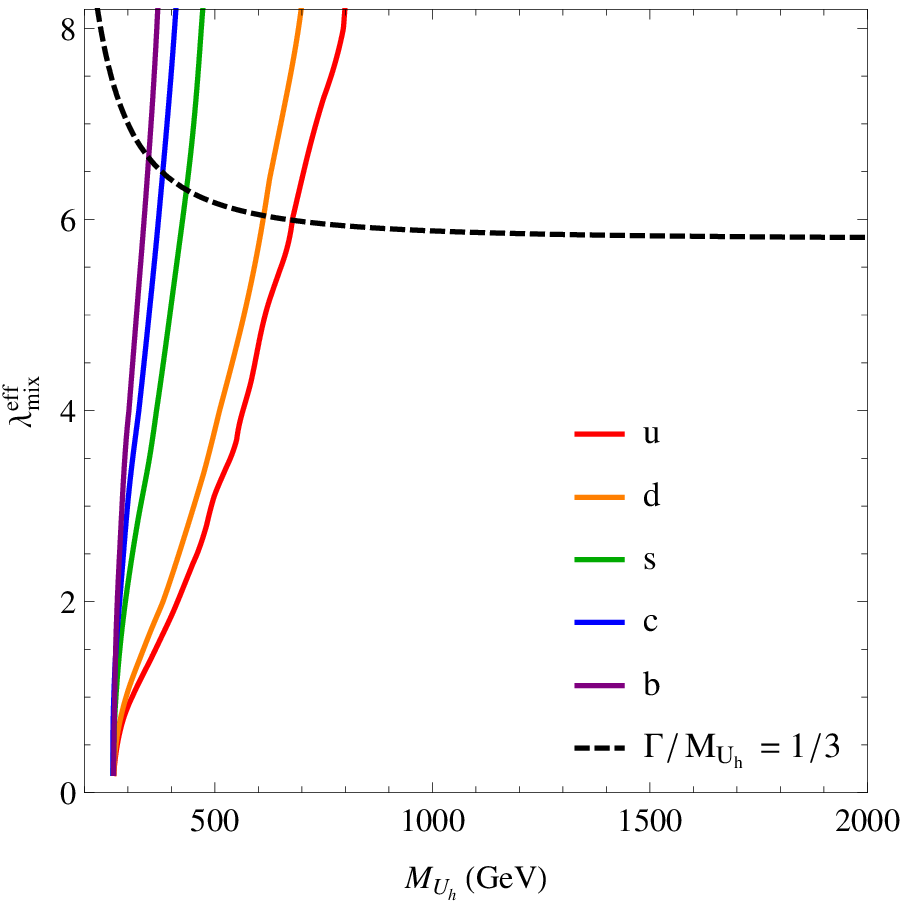} &
\includegraphics[scale=0.75]{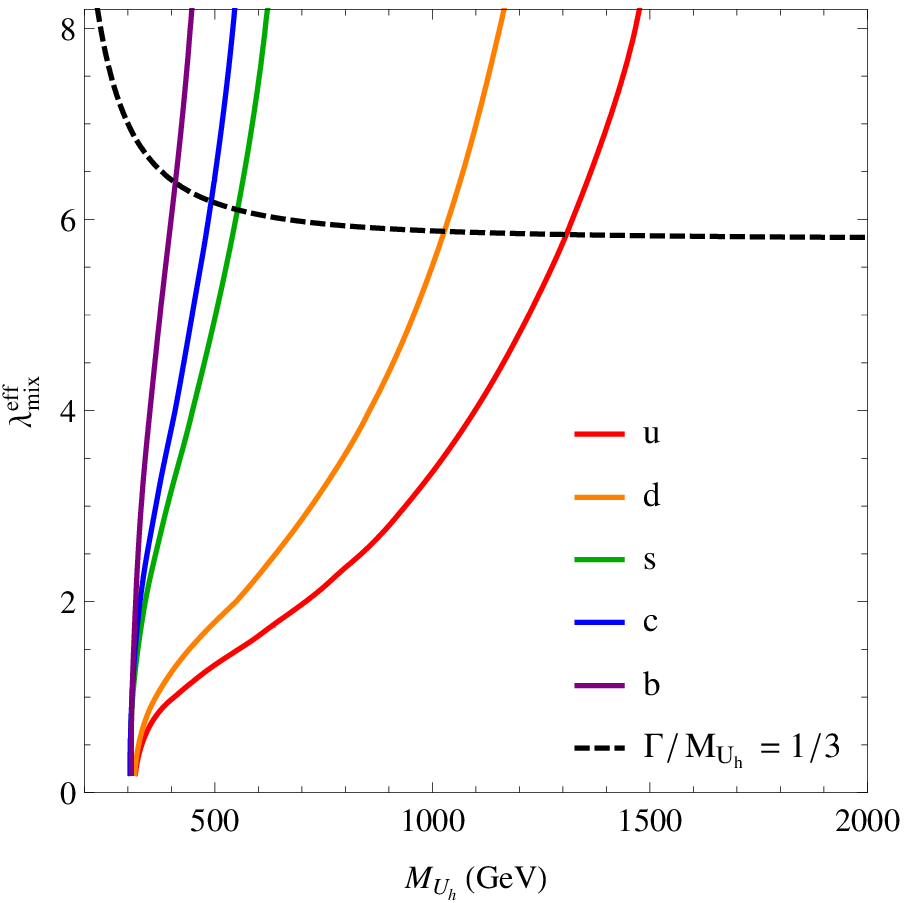}
\end{tabular}
\end{center}
\caption{The exclusion plots for partners of the $u,d,s,c$ and $b$ quark (curves from right to left) from $p_T^{\gamma\gamma}$ (upper left plot), $N_\mathrm{jets}$ (upper right plot), $p_T^{j_1}$ bins (lower left plot) and their combination (lower right plot). The parameter region to the top-left region from the respective curve is excluded at 95\% CL. For reference, the black dashed line in each of the plots shows the coupling above which the narrow-width-approximation does not apply anymore.}
\label{fig:pcresult}
\end{figure}

Fig.~\ref{fig:pcresult} shows the 95\% confidence level (CL) exclusion bounds on partners of partially composite light quarks  in the $\lambda^{\mathrm{\rm eff}}_{\mathrm{\rm mix}}$ vs. $M_{U_h}$ parameter space, which result from $p_T^{\gamma\gamma}$ bins, $p_T^{j_1}$ bins, $N_{\mathrm{jets}}$ bins, and the combined bound. For reference, the dashed line in each of the plots shows the coupling above which the width of the quark partner exceeds $1/3$ of its mass, so that the narrow-width-approximation is not applicable anymore.\footnote{This line is only given for reference. Our simulation and results do not rely on the narrow width approximation and remain valid in the full parameter space shown.} 

The dependence of the bound on $\lambda^{\mathrm{\rm eff}}_{\mathrm{\rm mix}}$ which is reflected in all three observables can be understood from the quark partner production cross sections shown in Figs.~\ref{fig:prodXsec} and \ref{fig:prodXsecsing}. For small $\lambda^{\mathrm{\rm eff}}_{\mathrm{\rm mix}}$, QCD pair production dominates for partners of all quark flavors. For non-suppressed $\lambda^{\mathrm{\rm eff}}_{\mathrm{\rm mix}}$, non-QCD single and pair production plays a role mainly for up- and down-quark partners for which the initial $qq$, $q\bar{q}$ and $\bar{q}\bar{q}$ are not substantially reduced by the PDFs. Parallel to the increased production of quark partners, the di-Higgs production due to $t$-channel exchange of a $u$ or $d$ quark partner is increased for the same reason (\cf Fig.~\ref{fig:prodXseczero}). Comparing the different observables, the strongest bound in our analysis arises from the $N_{\mathrm{jets}}\geq 3$ bin. The reason is that distribution of $p_T^{\gamma\gamma}$ (and also for $p_T^{j_1}$) for high $M_{U_h}$ masses is centered above the upper cutoff of the highest $p_T$ bin of $200 \gev$ ($140 \gev$), such that the majority of BSM events lies in the  $p_T^{\gamma\gamma}$ (or respectively $p_T^{j_1}$) overflow bin which we conservatively ignore in our main analysis. For an estimation of the constraints including the $p_T^{\gamma\gamma}$ and $p_T^{j_1}$ overflow bin we refer the interested reader to Appendix~\ref{app:overflow}.

As we have discussed before, the flavor-independent bound arises from small $\lambda^{\mathrm{\rm eff}}_{\mathrm{\rm mix}}$ region, in which QCD production dominates. Considering the all bins combined exclusion limit, the quarks partners are excluded up to
\begin{equation}
M_{U_h} \gtrsim
310 \;\mathrm{GeV} \mbox{ at 95\% CL.}
\end{equation}

\bigskip

The results in Fig.~\ref{fig:pcresult} are shown in terms of the effective mixing parameter $\lambda^{\mathrm{\rm eff}}_{\mathrm{\rm mix}}$ and the physical mass $M_{U_h}$ and apply to any model described by the effective Lagrangian Eq.~\eqref{Lpceff}. For concrete realizations of partially composite quark models, like the embedding of partners in the $\bf 5$ or $\bf 14$ of an $SO(5)$ multiplet, the parameters $\lambda^{\mathrm{\rm eff}}_{\mathrm{\rm mix}}$ and  $M_{U_h}$ are correlated, as discussed in Sec.~\ref{sec:PCtheory}. Obtaining a quark partner mass substantially below the composition scale $f$ implies an tight upper bound on $|\lambda^{\mathrm{\rm eff}}_{\mathrm{\rm mix}}|$ at this mass. For example, for a mass of $M_{U_h}=310 \gev$ and a compositeness scale $f=750 \gev$, using the expressions for the effective coupling in Table~\ref{table:pc} together with the expression for $M_{U_h}$ in Eq.~\eqref{pcmassh} yields  $|\lambda^{\mathrm{\rm eff}}_{\mathrm{\rm mix}}|\lesssim 0.07$ for a partner embedding into the $\bf 5$ and $|\lambda^{\mathrm{\rm eff}}_{\mathrm{\rm mix}}|\lesssim 0.18$ for a partner embedding into the $\bf 14$. For both these embeddings, the current bounds from Fig.~\ref{fig:pcresult} are thus dominated by QCD pair production and given by the flavor-independent bound.


\subsection{Results for fully composite light quarks}
As compared to partially composite quark models, the exclusion limit on fully composite models is weaker. The main reason is that for decays of partners of fully composite quarks, the branching fraction into a Higgs and a jet is only $\sim 25\%$. Furthermore, partners of fully composite quarks are only produced via QCD pair production. Quantitatively we find no significant signal excess over background in any of the $p_T^{\gamma\gamma}$ and $p_T^{j_1}$ bins. A $\chi^2$-fit to the $N_\mathrm{jets}$ data only results in a very weak bound of 
\begin{equation}
M_{U_h} \gtrsim
212 \;\mathrm{GeV}  \mbox{ at 95\% CL.}
\end{equation}


\section{Summary and Conclusions}\label{sec:conclusions}

In this article, we discussed the phenomenology of $SO(4)$ singlet partners of the $u$, $d$, $s$, $c$, and $b$ quark in minimal composite Higgs model. In Sec.~\ref{sec:models}, we derived the effective Lagrangian for light quark partners in a setup with partially composite right-handed quarks (\cf Eq.~\eqref{Lpceff}) and a setup with fully composite right-handed quarks (\cf Eq.~\eqref{Lfceff}).\footnote{The effective description for fully composite quarks assumes the absence of a set of four-fermion operators which are generically induced in composite models, but can be forbidden by an additional symmetry as discussed in Sec.~\ref{sec:FCHDops}.}  Based on the effective description, we discussed the dominating quark partner production channels and the single- and pair production cross section for the LHC at 8 and 14 TeV in Sec.~\ref{sec:pheno} (\cf Figs.~\ref{fig:prodXsec} and \ref{fig:prodXsecsing}). Both models predict an excess in di-Higgs production which represents a very promising discovery channel.
 In addition, partners of fully composite quarks also yield an excess in channels with two hard jets (or $b$-jets) and two weak gauge bosons. The implications of existing ATLAS and CMS searches for the non-di-Higgs final states are discussed in Sec.~\ref{sec:phenofc}. By these searches, only the mass of a fully composite bottom partner is constraint to $M_{U_h}>700 \gev$ \cite{CMS-12-019}.

In order to obtain the current LHC bounds on the other partners of fully composite as well as partially composite quarks, we used the ATLAS measurements of differential cross sections of the Higgs boson in the $h\rightarrow \gamma\gamma$ channel \cite{ATLAShiggs}. Amongst the distributions provided in Ref.~\cite{ATLAShiggs}, our simulation shows the largest deviations of signal as compared to data in the $p_T^{\gamma\gamma}$, the $N_{\mathrm{jets}}$ and the highest jet $p_T$ distribution. Our main results for partially composite quark models are displayed in Fig.~\ref{fig:pcresult} where we show the 95\% CL bound resulting from the afore mentioned distributions individually, as well as the combined result. As  a $\lambda^{\mathrm{\rm eff}}_{\mathrm{\rm mix}}$ and flavor independent bound on the singlet partner mass we obtained $M_{U_h}> 310\gev$. In models which allow for sizable $\lambda^{\mathrm{\rm eff}}_{\mathrm{\rm mix}}$, the mass bound is enhanced for up- and down-quark partners, while strange-, charm-, and bottom partners.\footnote{Note however, that typical partially composite models require small $\lambda^{\mathrm{\rm eff}}_{\mathrm{\rm mix}}$ if the partner quarks are substantially lighter than the composition scale $f$, as discussed in Secs.~\ref{sec:PCtheory} and \ref{sec:pcresults}.}
 For partners of fully composite quarks, we found a constraint of $ M_{U_h}> 212\gev$ at 95\% CL which is independent of flavor and $\lambda^{\mathrm{\rm eff}}_{\mathrm{\rm mix}}$. 

In order to determine the bounds discussed above, we conservatively did not include the signal excess in the $p_T^{\gamma\gamma}$ and $p_T^{j_1}$ overflow bins on which partial information is provided in Ref.~\cite{ATLAShiggs}. We refer the interested reader to Appendix~\ref{app:overflow} for the ``projected'' bounds which include the overflow bin data. Our study shows that one can improve the bounds presented here, once a dedicated high $p^\gamma\gamma_T$ Higgs search is available.

We also showed the need for dedicated experimental searches in the di-Higgs channel (with two, one or zero associated high $p_T$ jets or $b$-quarks) at the LHC, which will be the golden channel for exploring the relevant theory space for the $SO(4)$ singlet partners of partially composite quarks. Apart from direct di-Higgs searches, searches for a large number of high $p_T$ $b$-jets in the final state are promising to lead to an indirect constraint. Finally, we also expect that dedicated searches for boosted Higgs signals in the currently accumulated LHC data set can significantly improve the bound we derived in this article.


\bigskip

\emph{Acknowlegements:} This work was supported by the National Research Foundation of Korea(NRF) grant funded by the Korea government(MEST) (No. 2012R1A2A2A01045722),
and also supported by Basic Science Research Program through the National Research Foundation of Korea(NRF) funded by the ministry of Education,
Science and Technology (No. 2013R1A1A1062597). 

\bigskip

\textbf{Note added:} After submission of this article, first CMS results on di-Higgs searches (derived in the context of the channel $pp\rightarrow H\rightarrow hh$ in a two-Higgs-doublet model) became available \cite{CMSdihiggs}. To provide a naive estimate of the implications of these bounds for composite quark models, we compared the constraints on $\sigma\times BR$ given in  \cite{CMSdihiggs} to the cross sections in the composite Higgs models discussed in this article which results in $M_{U_h}\gtrsim 300 \gev$ for partially composite quark models and  $M_{U_h}\gtrsim 200 \gev$ for fully composite quark models.


\appendix


\section{Overflow bin analysis}\label{app:overflow}

\begin{figure}[h]
\begin{center}
\begin{tabular}{cc}
\includegraphics[scale=0.75]{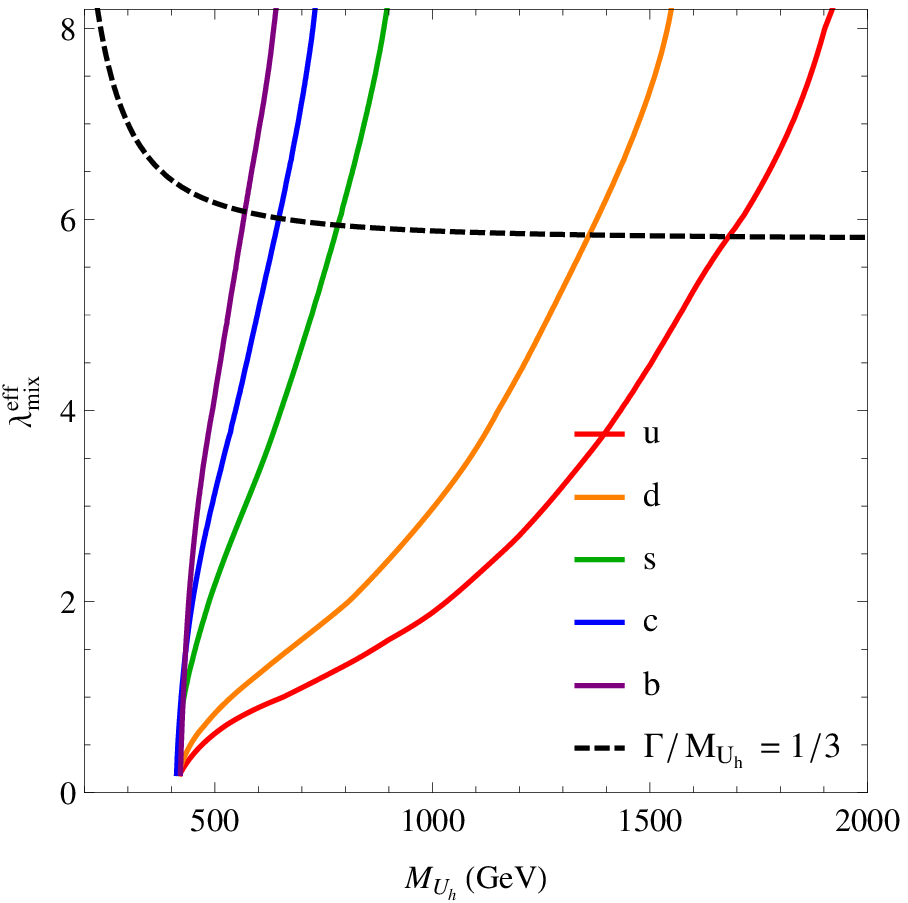} &
\includegraphics[scale=0.75]{Plots/Exclusion_NJet.eps}
\\
\includegraphics[scale=0.75]{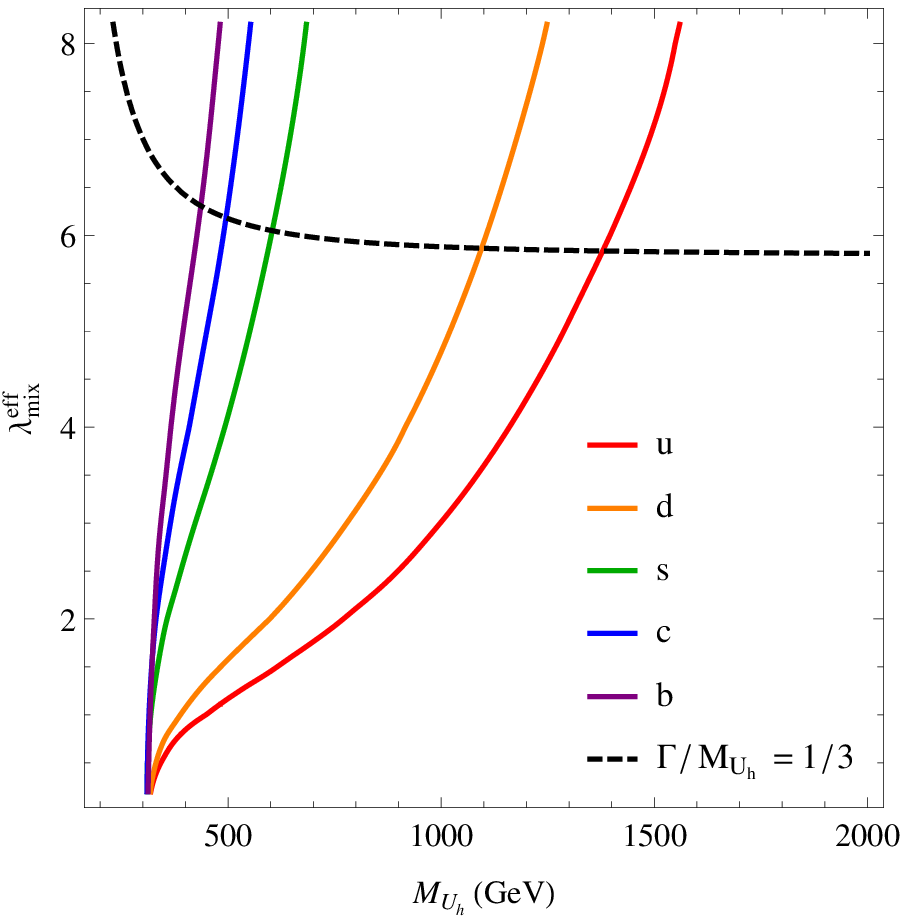} &
\includegraphics[scale=0.75]{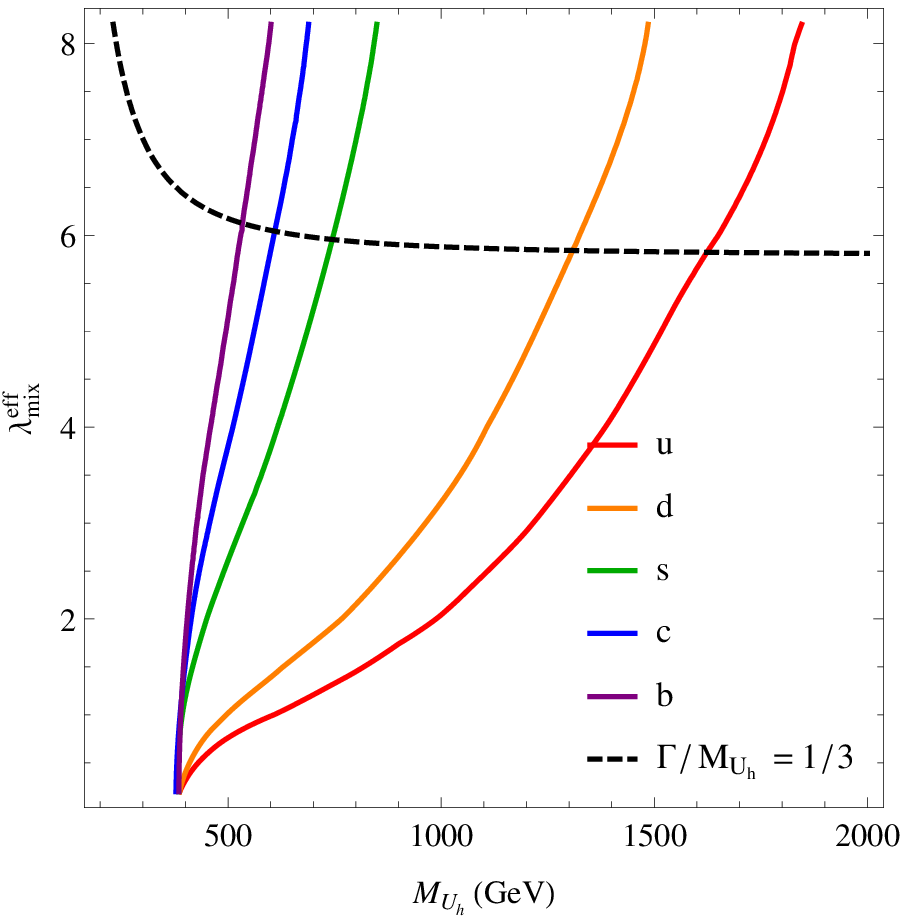}
\end{tabular}
\end{center}
\caption{Exclusion plots for partners of the $u,d,s,c$ and $b$ quark (from right to left). The plots show exclusions from  $p_T^{\gamma\gamma}$  (upper left), $N_\mathrm{jets}$ (upper right), $p_T^{j_1}$ (lower left) and their combination (lower right). Regions to the right of each curve are excluded by 95\% CL. To obtain these bounds, we included the $p_T^{\gamma\gamma}$ and $p_T^{j_1}$ overflow bin data in the analysis.}
\label{fig:pcresultOverflow}
\end{figure}

In  Sec.~\ref{sec:results} we conservatively did not include the  $p_T^{\gamma\gamma}$ and $p_T^{j_1}$ overflow bin data into our analysis. In this Appendix we repeat our analysis, including the overflow bins. Of course, the overflow bin data should be taken on a different footing., and bounds presented in this Appendix are to be taken with a grain of salt. The main purpose of this analysis is to demonstrate the impact of including high $p_T^{\gamma\gamma}$ and $p_T^{j_1}$ events in searches for Higgses from BSM events which -- like in our case -- originate from decays of BSM particles. 
 
The unfolding correction factor for overflow bins are not specified in Ref.~\cite{ATLAShiggs}. The specified correction factors for the other $p_T^{\gamma\gamma}$ bins vary between 150\% and 152\%, while the correction factors for $p_T^{j_1}$ vary between 138\% and 159\%. For our analysis we assumed the unfolding correction factor for overflow bins to be identical the correction factor of the highest $p_T$ bin of the respective distribution (150\% for $p_T^{\gamma\gamma}$ and 143\% for $p_T^{j_1}$). 
   
Fig.~\ref{fig:pcresultOverflow} shows the 95\% CL exclusion bounds on partners of partially composite light quarks in the $\lambda^{\mathrm{eff}}_{\mathrm{mix}}$ vs. $M_{U_h}$ parameter space, which result from the highest $p_T^{\gamma\gamma}$ bin, the highest $p_T^{j_1}$ bin, the $N_{\mathrm{jets}}\geq 3$, and the combined bound. The flavor and $\lambda^{\rm eff}_{\rm mix}$ independent bound arising from QCD pair produced quark partners is increased to 
\beq
M_{U_h} \gtrsim 385 \gev \mbox{ at 95\% CL.}
\eeq
As can be seen, the strongest constraint arises from the $p_T^{\gamma\gamma}$ overflow bin, followed by the $p_T^{j_1}$ overflow bin. In Sec.~\ref{sec:results} the dominant bound arose from the $N_{\mathrm{jets}}$ distribution. This demonstrates the impact of including the high $p_T^{\gamma\gamma}$ and $p_T^{j_1}$ events.

\bigskip

For fully composite quark models we find, that the bound is enhanced due to the  $p_T^{\gamma\gamma}$ overflow bin data while the $p_T^{j_1}$ overflow bin excess is too small to give an exclusion bound. Combining the constraints obtained from $p_T^{\gamma\gamma}$ and $N_\mathrm{jets}$ excesses we find a bound of 
\beq
M_{U_h} \gtrsim 240 \gev \mbox{ at 95\% CL.}
\eeq

\end{document}